\documentclass[preprint,preprintnumbers,showpacs,amssymb,amsmath,secnumarabic,tightenlines,nofootinbib,12pt]{revtex4}
\usepackage{bm}% bold math (revtex)
\usepackage{psfig}
\def\babar{\mbox{\sl B\hspace{-0.4em} {\scriptsize\sl A}\hspace{-0.4em} 
	B\hspace{-0.4em} {\scriptsize\sl A\hspace{-0.1em}R}}}
\def\fom{figure of merit}
\def\spr{StatPatternRecognition}
\long\def\simplex#1#2#3#4{
\begin{figure}[#1]
   \begin{center}
   \hbox{
   \quad
   \parbox[t]{14.5cm}{ \psfig{file=#2,width=12.5cm}
   \caption[c]{\small  \label{fig:#3} #4 } }
   }
   \quad
   \end{center}
\end{figure}
}
\long\def\duplex#1#2#3#4#5{
\begin{figure}[#1]
   \begin{center}
   \hbox{\hskip -0cm
   \quad 
   \parbox[t]{5.5cm}{ \psfig{figure=#2,width=8cm}
   } 
   \quad
   \parbox[t]{5.5cm}{ 
   \psfig{figure=#3,width=8cm} }
   }
   \caption[c]{\small \label{fig:#4} #5 }
   \quad
   \end{center} 
\end{figure}
}

\begin{document}

\preprint{}

\title{StatPatternRecognition: A C++ Package for Statistical Analysis of
	High Energy Physics Data
\footnote{Work partially supported by Department of Energy under Grant
DE-FG03-92-ER40701.}
}
\author{Ilya Narsky}
\email{narsky@hep.caltech.edu}
\affiliation{California Institute of Technology}
\date{\today}

\begin{abstract}
Modern analysis of high energy physics (HEP) data needs advanced
statistical tools to separate signal from background. A C++ package
has been implemented to provide such tools for the HEP community. The
package includes linear and quadratic discriminant analysis, decision
trees, bump hunting (PRIM), boosting (AdaBoost), bagging and random
forest algorithms, and interfaces to the standard backpropagation
neural net and radial basis function neural net implemented in the
Stuttgart Neural Network Simulator. Supplemental tools such as
bootstrap, estimation of data moments, and a test of zero correlation
between two variables with a joint elliptical distribution are also
provided. The package offers a convenient set of tools for imposing
requirements on input data and displaying output. Integrated in the
\babar\ computing environment, the package maintains a minimal set of
external dependencies and therefore can be easily adapted to any other
environment. It has been tested on many idealistic and realistic
examples.
\end{abstract}


\maketitle

\section{Introduction}

Pattern recognition, or pattern classification, developed
significantly in the 60's and 70's as an interdisciplinary
effort. Several comprehensive reviews of the field are
available~\cite{htf,webb,haykin,kuncheva}.

The goal of pattern classification is to separate events of different
categories, e.g., signal and background, mixed in one data
sample. Using event input coordinates, a classifier labels each event
as consistent with either signal or background hypotheses. This
labeling can be either discrete or continuous. In the discrete case,
with each category represented by an integer label, the classifier
gives a final decision with respect to the event category. In case of
continuous labeling, the classifier gives real-valued degrees of
consistency between this event and each category. In the simplest case
of only two categories, a continuous classifier produces one real
number for each event with high values of this number corresponding to
signal-like events and low values corresponding to background-like
events. The analyst can then impose a requirement on the continuous
classifier output to assign each event to one of the two categories.

A primitive pattern classifier with discrete output is a binary split
in one input variable. This classifier has been used by every
physicist many times and is typically referred to as ``cut'' in HEP
jargon. More advanced classifiers are also available. Linear
discriminant analysis was introduced by Fisher~\cite{fisher} in 1936
and became a popular tool in analysis of HEP data. Neural
networks~\cite{nn}, originally motivated by studies of the human
brain, were adopted by HEP researchers in the late 80's and early
90's~\cite{nnhep}. The feedforward backpropagation neural net with a
sigmoid activation function appears to be the most popular
sophisticated classifier used in HEP analysis today.

The range of available classifiers is, of course, not limited to the
Fisher discriminant and neural net. Unfortunately, classification
methods broadly used by other communities are largely unknown to
physicists. Radial basis function networks~\cite{rbf} are seldom used
in HEP practice~\cite{rbfinhep}. Decision trees~\cite{cart,c45}
introduced in the 80's have been barely explored by
physicists~\cite{treesinhep}. There is only one documented application
of boosted decision trees~\cite{boosted}, a powerful and flexible
classification tool, to analysis of HEP data~\cite{roe}. Multivariate
adaptive regression splines~\cite{mars}, projection
pursuit~\cite{pursuit} and bump hunting~\cite{prim} have yet to find
their way to HEP analysts.

To some extent, these methods can be popularized through active
advertisement. However, comparison of various classifiers on the same
input data and therefore an educated choice of an optimal classifier
for the problem at hand are hardly possible without consistent and
reliable code. In principle, implementations of all classification
methods listed above are available, as either commercial or free
software. In practice, one has to deal with different formats of input
and output data, different programming languages, different levels of
support and documentation, and sometimes with incomprehensible
coding. Until an average analyst finds a way to feed data to various
classifiers with minimal effort, advanced pattern recognition in HEP
will be left to a few enthusiasts.

\spr\ is an attempt to provide such consistent code
for physicists. This package implements several classifiers suited for
physics analysis and serves them together in a flexible framework.

%Detailed instructions on the software use are included in the
%package. This note gives a short summary of implemented methods,
%discusses their performance on examples and briefly describes package
%capabilities.

This note describes basic mathematical formalism behind the
implemented classifiers, illustrates their performance with examples
and gives general recommendations for their use. If you are planning
to apply classification methods to practice, I strongly recommend that
you look more deeply into the topic. The comprehensive
reviews~\cite{htf,webb,haykin,kuncheva} are an excellent place to
start.

Section~\ref{sec:lqda} describes linear and quadratic discriminant
analysis broadly known in the physics community as
``Fisher''. Section~\ref{sec:nn} covers the feedforward
backpropagation neural net and the radial basis function neural net,
two classifiers implemented in the Stuttgart package. Although \spr\
does not offer tools for training neural nets and can be used only for
reading trained network configurations and computing network response
for input data, I felt obligated to include a brief description of
the formalism for these two methods as well. Section~\ref{sec:trees}
describes the \spr\ implementation of decision
trees. Section~\ref{sec:hunter} describes PRIM, an algorithm for bump
hunting. Section~\ref{sec:ada} describes our implementation of
AdaBoost, a quick, powerful, and robust
classifier. Section~\ref{sec:combiner} discusses a method for
combining classifiers implemented in \spr. A description of other
tools useful for classification and included in the package can be found
in Section~\ref{sec:other}. Technical details of the C++
implementation are briefly reviewed in Section~\ref{sec:cpp}. Finally,
application of several classifiers to a search for the radiative
lepton decay $B\to\gamma l\nu$ at \babar\ is discussed in
Section~\ref{sec:lnugamma}.

\section{Classifiers}

The main feature of a classifier is its predictive power, i.e., how
efficiently this classifier separates events of different categories
from each other. Mathematical definition of the predictive power can
vary from analysis to analysis. For example, in some situations one
might want to minimize the fraction of misclassified events, while
under other circumstances one might focus on maximizing the signal
significance, $S/\sqrt{S+B}$, where $S$ and $B$ are signal and
background, respectively, found in the signal region. Irrespective of
the exact mathematical definition, non-linear classifiers such as
neural nets or boosted decision trees typically provide a better
predictive power than simple robust classifiers such as binary splits
or the linear Fisher discriminant.

For optimization and estimation of the predictive power of a
classifier, a three-stage procedure is typically used: training,
validation, and test. First, the classifier is optimized on a training
data set. The classifier performance on a validation set is used to
stop the training when necessary. Finally, the predictive power of the
classifier is estimated using test data. The three data sets must be
independent of each other. Non-compliance with this three-stage
procedure can result in sub-optimal classifier performance and a
biased estimate of the predictive power. For example, the
classification error tends to decrease for the training sample even
after the error reached minimum and turned over for the validation
sample. This phenomenon is called ``overtraining''. Estimates of the
predictive power obtained from training or validation sets are most
usually over-optimistic.  These considerations are less important for
robust classifiers. For instance, a simple binary split in one variable
does not need validation, and the efficiency of the imposed cut for
the training set is usually very close to that for the test set for
large data samples. For small samples or non-linear classifiers, the
three-stage routine is a necessity.

Other important characteristics of classifiers include
interpretability, ability to deal with irrelevant inputs, stability of
training and performance in high-dimensional data, ease of training,
and classification response time per event. An ideal classifier offers
a high predictive power and provides insight into the data
structure. These two requirements are unfortunately at odds with each
other. Powerful classifiers such as neural nets or boosted decision
trees typically work as black boxes. On the other hand, decision trees
split data into rectangular regions suitable for easy interpretation
but rarely provide a competitive quality of classification. This
motivated Breiman to postulate his uncertainty
principle~\cite{uncertainty_principle}:
\begin{equation*}
\mbox{Ac}\cdot\mbox{Sm} > b\ ,
\end{equation*}
where $\mbox{Ac}$ stands for accuracy, $\mbox{Sm}$ stands for
simplicity, and $b$ is the Breiman constant\footnote{In case you were
fooled, this is an example of statistician's humor.}. An optimal
classifier for every problem has to be chosen with regard to all its
features.

Methods implemented in \spr\ can only process data sets with two event
categories --- signal and background. Perhaps, one day I will extend
these implementations to include multi-category classification --- if
the community shows sufficient interest for such multi-class methods.

\subsection{Linear and Quadratic Discriminant Analysis}
\label{sec:lqda}

If the likelihood function for each class is a multivariate Gaussian
\begin{equation}
f_c(x) = \frac{W_c}{ (2\pi)^{d/2} |\Sigma_c|^{1/2} }
   \exp \left[ -\frac{1}{2} (x-\mu_c)^T \Sigma_c^{-1} (x-\mu_c) \right];
	\ \ \ c=0,1;
\end{equation}
the two classes can be separated by taking the log-ratio of the
two likelihoods:
\begin{eqnarray}
\label{eq:qda}
\log \frac{f_1(x)}{f_0(x)} =
	\log \left( \frac{W_1}{W_0} \right) -
	\frac{1}{2} \log \left( \frac{|\Sigma_1|}{|\Sigma_0|} \right) -
	\frac{1}{2} 
	\left( \mu_1^T\Sigma_1^{-1}\mu_1 - \mu_0^T\Sigma_0^{-1}\mu_0 \right)
	\nonumber \\
	+\ x^T \left( \Sigma_1^{-1}\mu_1 - \Sigma_0^{-1}\mu_0 \right) -
	\frac{1}{2} x^T \left( \Sigma_1^{-1} - \Sigma_0^{-1} \right) x\ .
\end{eqnarray}
Above, $x$ is a $d$-dimensional vector of point coordinates, $\mu_c$
is a $d$-dimensional mean vector for class $c$, $\Sigma_c$ is a
$d\times d$ covariance matrix for class $c$ with determinant
$|\Sigma_c|$, and $W_c$ is the total weight of events in class
$c$. Here and below I use index 0 for background and 1 for signal.  In
reality, the mean vectors and covariance matrices are unknown and
need to be estimated from observed data:
\begin{equation}
\hat{\mu}_c = \frac{ \sum_{n=1}^{N_c} w_n^{(c)} x_n^{(c)} }
	{ \sum_{n=1}^{N_c} w_n^{(c)} }; \ \ \ \ \ 
\hat{\Sigma}_c = 
\frac{ \sum_{n=1}^{N_c} w_n^{(c)} 
	(x_n^{(c)}-\hat{\mu}_c) (x_n^{(c)}-\hat{\mu}_c)^T }
	{ \sum_{n=1}^{N_c} w_n^{(c)} };
\end{equation}
where $x_n^{(c)}$ is the observed vector for event $n$ in class $c$,
and $w_n^{(c)}$ is the weight of this event: $\sum_{n=1}^{N_c}
w_n^{(c)} = W_c$. Signal and background can now be separated by
requiring the log-ratio~(\ref{eq:qda}) to be above a certain value.

If the covariance matrices of the two classes are equal, this
expression simplifies to
\begin{equation}
\label{eq:fisher}
\log \frac{f_1(x)}{f_0(x)} =
	\log \left( \frac{W_1}{W_0} \right) -
	\frac{1}{2} (\mu_1-\mu_0)^T \Sigma^{-1} (\mu_1+\mu_0) 
	+ x^T \Sigma^{-1} (\mu_1-\mu_0) \ ,
\end{equation}
where $\Sigma$ is the common covariance matrix usually estimated as
\begin{equation}
\hat{\Sigma} = \frac{W_0\hat{\Sigma}_0+W_1\hat{\Sigma}_1}{W_0+W_1}\ .
\end{equation}
Formula~(\ref{eq:fisher}) is widely known among physicists as ``the
Fisher discriminant''; in the statistics literature this method is
usually referred to as ``linear discriminant analysis'' (LDA). If the
covariance matrices are not equal, the quadratic term in
Eqn.~(\ref{eq:qda}) gives rise to quadratic discriminant analysis
(QDA).

The three constant terms in the expression~(\ref{eq:qda}), often
omitted by analysts, are needed for proper normalization. With these
terms included, the equality $\log(f_1(x)/f_0(x))=0$ implies that the
likelihoods for the two classes at point $x$ are equal. The region
$\log(f_1(x)/f_0(x))>0$ is therefore populated with signal-like events
and the region $\log(f_1(x)/f_0(x))<0$ is populated with
background-like events.

If the signal and background densities are truly Gaussian with equal
covariance matrices, the linear discriminant~(\ref{eq:fisher}) for each
class has a univariate Gaussian distribution with mean
$\log(W_1/W_0)+\Delta/2$ (signal) or $\log(W_1/W_0)-\Delta/2$
(background) and standard deviation $\sqrt{\Delta}$, where
$\Delta=(\mu_1-\mu_0)^T\Sigma^{-1}(\mu_1-\mu_0)$. The overall
distribution of the discriminant~(\ref{eq:fisher}) is therefore a sum of
two Gaussians with areas $W_1$ and $W_0$. The separation, in numbers
of sigmas, between the two univariate Gaussians is
$N_\sigma=\sqrt{\Delta/2}$.

If the true signal and background densities are indeed Gaussian and if
the data have enough points to obtain reliable estimates of their
covariance matrices, the quadratic discriminant will perform at least
not worse than the linear one and usually better. In reality, the
signal-background separation is often improved by keeping the
quadratic term in Eqn.~(\ref{eq:qda}); however, if one of the above
conditions is not satisfied, omission of the quadratic term can
occasionally result in a better predictive power.

The quadratic discriminant can give a substantial improvement over the
linear discriminant, for instance, if the two densities are centered
at the same location $\mu_0=\mu_1$ but one of the densities is more
spread than the other: $|\Sigma_0|>|\Sigma_1|$. The linear
discriminant is useless in this case. The quadratic discriminant
separates signal from background by drawing a 2nd-order surface
$x^T(\Sigma_1^{-1}-\Sigma_0^{-1})x=\mbox{const}$. If the matrix
$\Sigma_1^{-1} - \Sigma_0^{-1}$ is positive definite, this surface is
elliptical, wrapped around the more compact density, as shown in
Fig.~\ref{fig:qda_gauss_and_uniform}. For pedagogical reasons, I show
in Fig.~\ref{fig:lin_gauss_and_uniform} an attempt to classify the
same data using a linear Fisher discriminant.

\simplex{htbp}{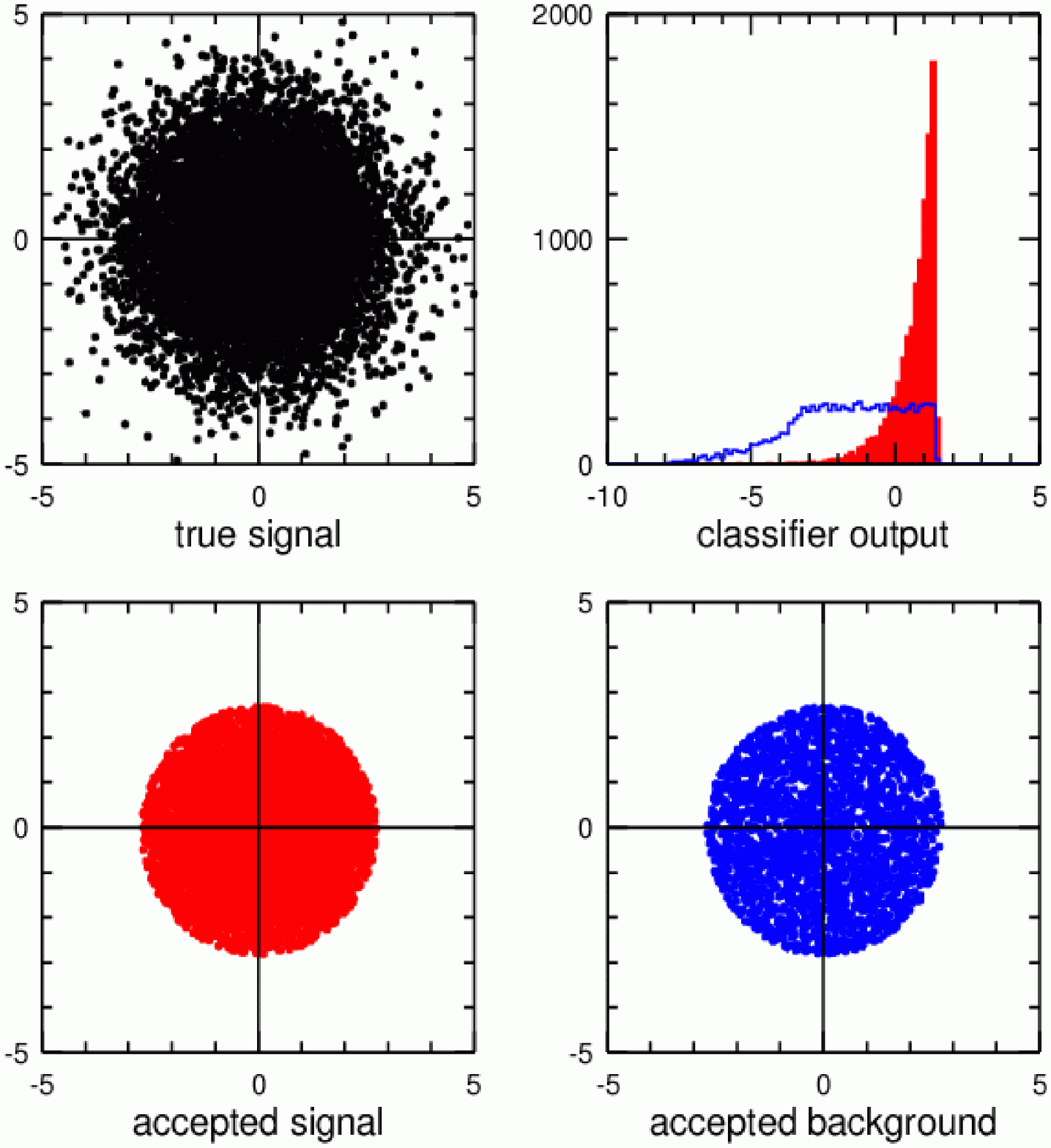}{qda_gauss_and_uniform}{
Separation of a signal bivariate Gaussian with mean $(0,0)$ and
identity covariance matrix from uniform background by quadratic
discriminant analysis. 10,000 true signal events (top left). QDA
output for signal events (solid red) and background events (hollow
blue) is shown in the upper right plot. True signal events selected by
requiring the QDA output to be positive (bottom left). True background
events selected by the same requirement (bottom right).}

\simplex{htbp}{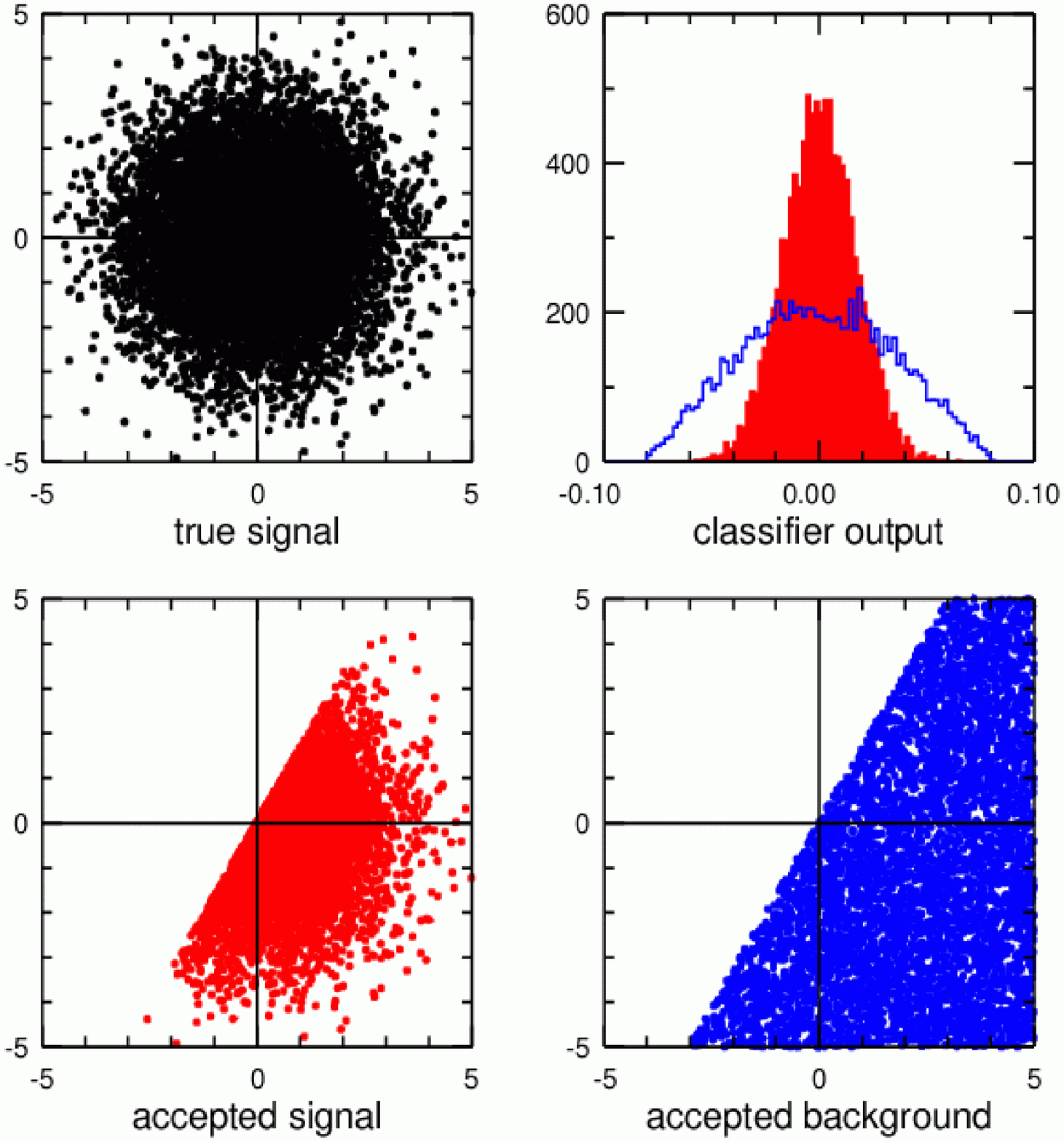}{lin_gauss_and_uniform}{
Separation of a signal bivariate Gaussian with mean $(0,0)$ and
identity covariance matrix from uniform background by linear
discriminant analysis. 10,000 true signal events (top left). LDA
output for signal events (solid red) and background events (hollow
blue) is shown in the upper right plot. True signal events selected by
requiring the LDA output to be positive (bottom left). True background
events selected by the same requirement (bottom right).}

\subsection{Neural Networks}
\label{sec:nn}

The Stuttgart Neural Network Simulator (SNNS)~\cite{snns} provides a
variety of tools for classification including feedforward neural
networks with backpropagation and radial basis function
networks. Because the SNNS package is well maintained and has already
earned a good reputation in the world of HEP, there is no need to
reimplement these methods. However, SNNS does not offer convenient
tools for plotting its output and classifying new data with a trained
network configuration. \spr\ implements interfaces to the two types of
networks mentioned above. The user can read a saved SNNS network
configuration and apply it to any independently-supplied data. The
format of input ascii files implemented in \spr\ is
very close to that used by SNNS. The user can therefore switch between
SNNS and \spr\ with minimal effort.

A disadvantage of the SNNS package is its inability to process
weighted events. All other methods implemented in
\spr\ accept weighted data.

\subsubsection{Feedforward Neural Network with Backpropagation}
\label{sec:stdnn}

A feedforward neural network consists of one input, one output and
several hidden layers linked sequentially. Training of a neural net
consists of two stages: forward and backward propagation. At the first
stage, a $d$-dimensional vector $x$ of input event coordinates
$z_i^{(0)}=x_i;\ 1\leq i\leq d;$ is propagated from the input to the
output layer using
\begin{equation}
\begin{array}{rcl}
v_i^{(l+1)} & = & \sum_{j=1}^{d^{(l)}} \alpha_{ji}^{(l)} z_j^{(l)};\\
z_i^{(l+1)} & = & {\mathcal A}_i^{(l+1)} \left( v_i^{(l+1)} \right);
\end{array}
\end{equation}
where $z_i^{(l)}$ is the output generated by the $i$th node in layer
$l$, $d^{(l)}$ is the number of nodes in layer $l$,
$\alpha_{ji}^{(l)}$ is a linear weight associated with the link
between node $j$ in layer $l$ and node $i$ in layer $l+1$,
$v_i^{(l+1)}$ is the local field, and ${\mathcal A}_i^{(l)}(v)$ is an
activation function for node $i$ in layer $l$. The sequential order of
this propagation from input to output through a predetermined set of
links explains why such a network is called ``feedforward'' or
``acyclic''. For a two-category classification problem, the output
layer has only one node with real-valued output ranging from 0 to 1.
The real-valued output $o=z_1^{(L)};\ d^{(L)}=1;$ of a network with
$L+1$ layers indexed from 0 to $L$ is used to compute the quadratic
classification error for each training event
\begin{equation}
\label{eq:quadr_loss}
{\mathcal E} = \left( y-o \right)^2\ ,
\end{equation}
where $y$ is the true class of the input vector $x$, 0 for background
and 1 for signal.

The network weights $\alpha_{ij}^{(l)}$ are then updated by
propagating the computed classification error backwards from the
output to the input layer:
\begin{equation}
\begin{array}{rcl}
\Delta \alpha_{ij}^{(l)} & = & \eta \delta_j^{(l+1)} z_i^{(l)};\\
\delta_i^{(l)} & = & \left. \frac{d{\mathcal A}_i^{(l)}}{dv}
	\right|_{ v=v_i^{(l)} }
	\sum_{j=1}^{d^{(l+1)}} \alpha_{ij}^{(l)} \delta_j^{(l+1)};
\end{array}
\end{equation}
where $\Delta \alpha_{ij}^{(l)}$ are weight adjustments,
$\delta_i^{(l)}$ is the local gradient at node $i$ of layer $l$, and
$\eta$ is the learning rate of the network. At the output node $l=L$
the local gradient is simply
\begin{equation}
\delta_1^{(L)} = ( y-o ) \left. \frac{d{\mathcal A}_1^{(L)}}{dv}
	\right|_{ v=v_1^{(L)} }\ .
\end{equation}

This two-pass procedure is repeated for each training event. After the
pool of training events is exhausted, the neural net goes back to the
first event and reprocesses the whole training sample. 

For the activation ${\mathcal A}$, a sigmoid, or logistic, function is
often chosen:
\begin{equation}
{\mathcal A}(v) = \frac{1}{1+\exp(-av)};\ a>0.
\end{equation}

The backpropagation algorithm described above and the sigmoid
activation function are among the many possibilities offered by the
SNNS package. They are emphasized here simply because they are often
chosen by HEP analysts.

To apply a neural net to classification, the user must choose the
number and size of hidden layers and specify the learning rate
$\eta$. Both selections are typically optimized by studying the
network performance on training and validation data.

Compared to other classifiers, neural networks usually offer an
excellent predictive power. Their main disadvantage is the lack of a
meaningful interpretation. A neural net works as a black box. Its
output is a non-linear function which is hard to visualize, especially
in high dimensions.

Training of neural nets, however, is subject to certain subtleties. A
neural net can get ``confused'' if it is presented with strongly
correlated input variables or irrelevant variables that do not
contribute to the predictive power. Mixed input data, i.e., a
combination of continuous and discrete variables, can also cause
training instabilities. Adding extra variables to a neural net not
only increases the size of the input layer but also requires an
expansion of hidden layers for an adequate representation of the data
complexity. In many dimensions, therefore, training of neural nets can
be painfully slow.

A neural net can fail even in low dimensions if it is presented with
sufficiently complex data. Consider, for example, separating two
signal Gaussians from uniform background in two dimensions, as shown
in Fig.~\ref{fig:twogauss_and_uniform}. A 2:4:2:1 fully-connected
feedforward neural net with a sigmoid activation function efficiently
finds one Gaussian but not the other. The network performance can be,
of course, improved. For example, one could build two independent
networks, each optimized for one of the two Gaussians, and then
combine them into one common network. But to perform such an exercise,
one must already have a good knowledge of the data structure. The
neural net construction would be therefore unnecessary --- if we
already know so much about the data, we can separate signal from
background ``by hand'', i.e., using simple and well-understood
selection requirements.

\simplex{htbp}{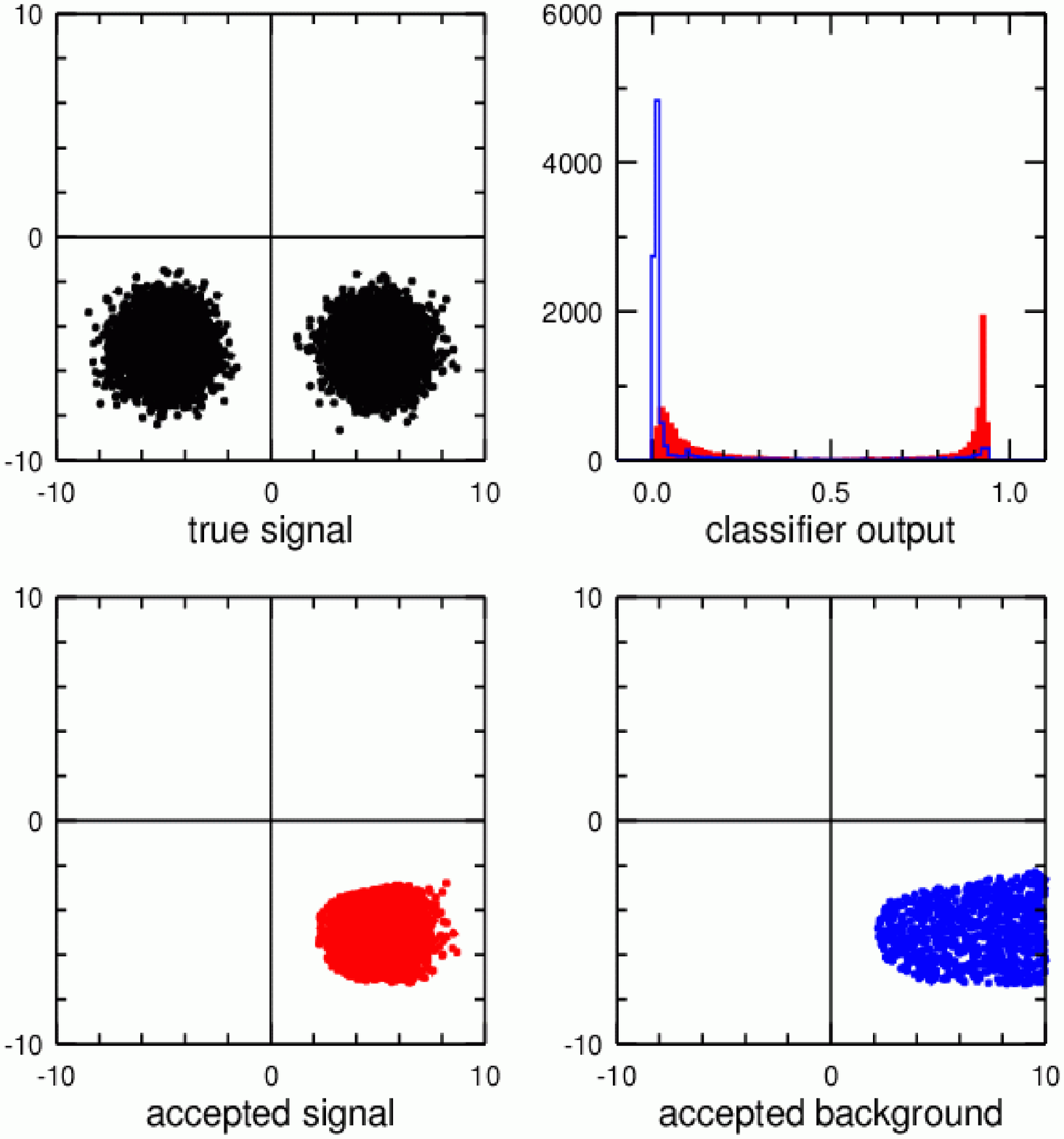}{twogauss_and_uniform}{
Separation of two bivariate Gaussians from uniform background by a
2:4:2:1 feedforward backpropagation neural net with a sigmoid
activation function. Two signal Gaussians, 5000 events each, with
means $(-5,-5)$ and $(5,-5)$, respectively, and identity covariance
matrices (top left). Neural net output for signal (solid red) and
background (hollow blue) events (top right). True signal events
selected by requiring the neural net output to be above 0.5 (bottom
left). True background events selected by the same requirement (bottom
right). The background density, not shown here, is uniform on the
square $(-10,-10)-(10,10)$. In this example, the neural net fails to
find the left signal Gaussian.}

\subsubsection{Radial Basis Function Networks}

In the radial basis function (RBF) formalism~\cite{rbf}, an unknown
real-valued function $f(x)$ is regressed on an input space $x$ using
\begin{equation}
\label{eq:rbf_regression}
f(x) = \sum_{m=1}^M \alpha_m K(x,t_m)\ ,
\end{equation}
where $\{t_m\}_{m=1}^M$ are regression centers and $K(x,t)$ is the
regression kernel. The regression coefficients $\alpha_m$ can be
estimated from observed data $\{x_n,y_n\}_{n=1}^N$:
\begin{equation}
\label{eq:rbf_weights}
\alpha = (G^TG+\lambda T)^{-1} G^T y\ ,
\end{equation}
where $\alpha$ is an $M$-dimensional vector of coefficients, $y$ is an
$N$-dimensional vector of the function values observed at points
$\{x_n\}_{n=1}^N$, $G$ is an $N\times M$ matrix defined by
$G_{nm}=K(x_n,t_m)$, $T$ is an $M\times M$ matrix defined by
$T_{ij}=K(t_i,t_j)$, and $\lambda$ is a regularization parameter
included to avoid overtraining. The kernel is usually chosen to be
radially symmetric,
\begin{equation}
K(x_n,t_m) = K_{\mbox{rbf}}(|x_n-t_m|)\ ,
\end{equation}
hence the name ``radial basis function''. The two-class recognition
problem is equivalent to the regression problem with the real-valued
function $f(x)$ replaced by an integer-valued function $f(x)$ allowed
to take only two values, 0 for background and 1 for signal.

Three popular choices for the kernel $K_{\mbox{rbf}}(r)$ have been
implemented in the Stuttgart package~\cite{snns}:
\begin{itemize}
	\item Gaussian
\begin{equation}
K_{\mbox{rbf}}(r) = \exp \left( -\frac{r^2}{\sigma^2} \right);
\end{equation}

	\item Multiquadratic
\begin{equation}
K_{\mbox{rbf}}(r) = \sqrt{\sigma^2 + r^2};
\end{equation}

	\item Thin plate spline
\begin{equation}
K_{\mbox{rbf}}(r) = \left( \frac{r}{\sigma} \right)^2 
	\log \left( \frac{r}{\sigma} \right);
\end{equation}
\end{itemize}
where $\sigma$ is a scale parameter.

The RBF formalism is represented by a 3-layer neural network. The
input layer consists of $d$ nodes, one for each component of the
$d$-dimensional input vector $x$. $M$ nodes representing the
regression centers compose the hidden layer. The output layer for a
two-class recognition problem is a single node with real-valued output
ranging from 0 to 1. Components of the input vector $x$ are delivered
to the hidden nodes which compute distances between the input vector
and the corresponding regression centers. The hidden units are then
activated by applying the chosen kernel to the computed distances. The
result of the activation is linearly propagated from the hidden layer
to the output node. This node applies an activation function, either
an identity or sigmoid transformation, to its input to produce the
network output. Shortcut connections between the input and output
layers are allowed but not required; such shortcuts are implemented by
adding linear terms to the regression
formula~(\ref{eq:rbf_regression}). One can also bias the output node
by adding a constant term to its input. The final expression for the
network output is then
\begin{equation}
o(x) = {\mathcal A} \left( \sum_{m=1}^M \alpha_m K_{\mbox{rbf}}(|x-t_m|)
	+ \beta x + b \right)\ ,
\end{equation}
where ${\mathcal A}$ is the activation function for the output node,
$\beta$ is a $d$-dimensional vector of the linear shortcut
coefficients, and $b$ is the bias applied to the output node.

Several algorithms are available for training RBF networks; for a
brief survey see a subsection on learning strategies for RBF nets in
Ref.~\cite{haykin}. The SNNS implementation~\cite{snns} minimizes
quadratic error
\begin{equation}
{\mathcal E} = \sum_{n=1}^N \left( y_n-o(x_n) \right)^2
\end{equation}
by using gradient descent. In this approach, the weights $\alpha_m$
are not computed using Eqn.~(\ref{eq:rbf_weights}) but adjusted to
minimize the classification error on training data. The regularization
parameter $\lambda$ is not used, and smootheness of the regression
surface is maintained by stopping the training routine when the
validation error reaches minimum.

A non-trivial part of the training procedure is initialization of the
regression centers $\{t_m\}_{m=1}^M$. A conservative approach is to
treat every observed point as a regression center by setting $M=N;\
t_n=x_n;\ 1\leq n\leq N$. While this approach generally guarantees a
high predictive power, it is hardly practical for large data sets. In
practice, one usually selects a smaller number of regression centers
and then samples the observed distribution to find an efficient
initial center assignment. This sampling can be accomplished by a
number of algorithms; see a section on RBF implementation in the SNNS
manual~\cite{snns}.

RBF nets share most flaws and advantages with the backpropagation
neural network of Section~\ref{sec:stdnn}.

The SNNS implementation of the RBF training procedure optimizes the
RBF weights $\alpha$, the scale parameter $\sigma$ for each RBF center
and positions of the RBF centers $\{t_m\}_{m=1}^M$. The RBF training
mechanism is therefore quite flexible and provides a good predictive
power. However, the large number of optimized parameters makes an RBF
network more fragile than a backpropagation neural net of the same
size. RBF networks also suffer a great deal from the ``curse of
dimensionality''. In high-dimensional data, the performance of
distance-based classifiers crucially depends on the chosen definition
of distance between two points. In addition, the number of points
required to adequately sample a distribution grows exponentially with
the dimensionality of input space thus making every realistic data set
undersampled in a high-dimensional representation. The combination of
these two features makes every distance-based classifier highly
unstable in a multidimensional setting.

Despite these setbacks, RBF nets should be considered as a good
competitor to the backpropagation neural net in low dimensions. For
example, the problem with two signal Gaussians and uniform background
discussed in Section~\ref{sec:stdnn} can be solved, as shown in
Fig.~\ref{fig:rbf_twogauss_and_uniform}, using a very simple RBF
network with only 3 regression centers --- one for each Gaussian and
one for background. If the two signal peaks were not Gaussian, more
regression centers would be needed. But it is clear that the RBF net
is superior over the standard backpropagation neural net in this case.

\simplex{htbp}{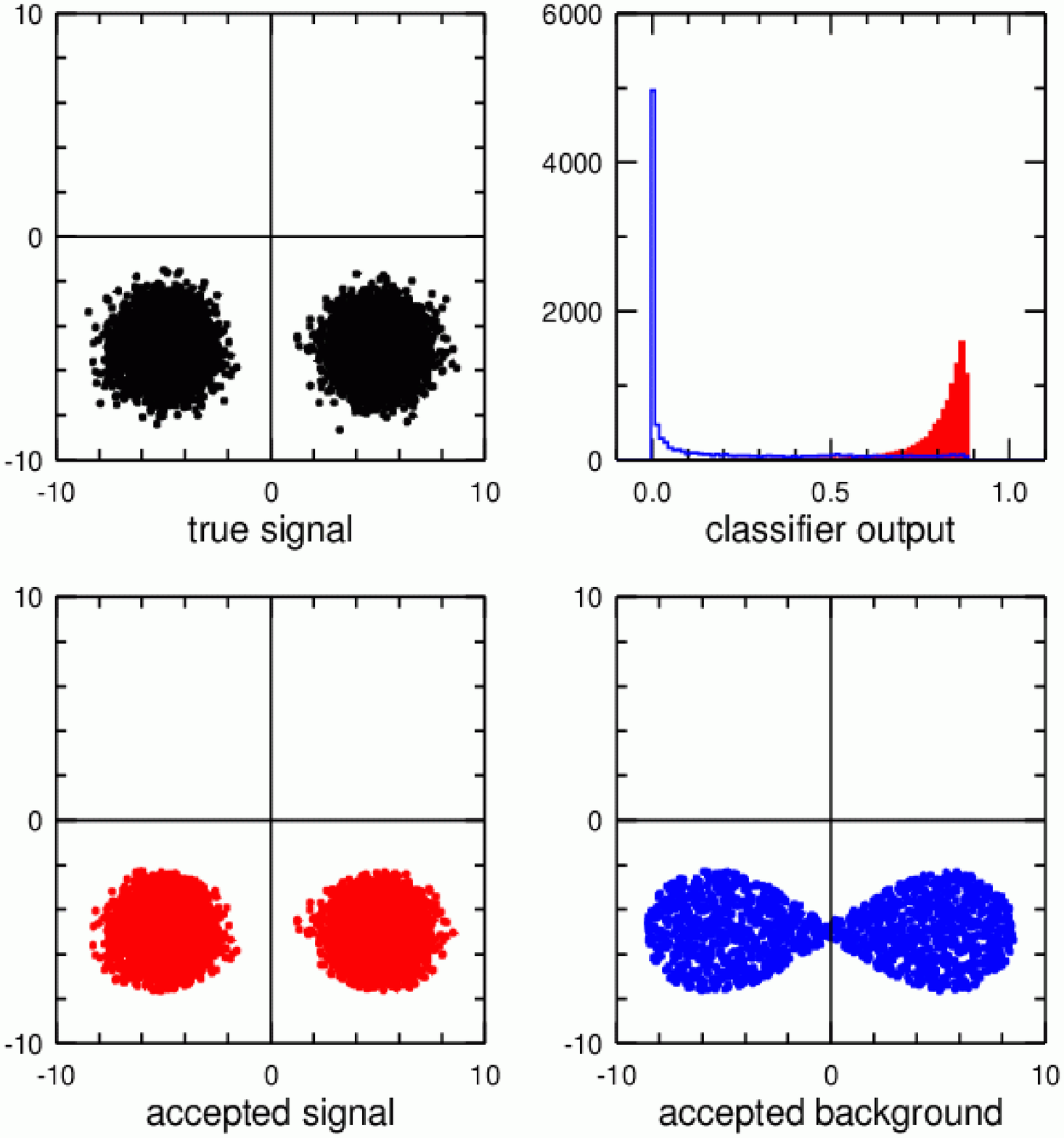}{rbf_twogauss_and_uniform}{
Separation of two bivariate Gaussians from uniform background by a
2:3:1 RBF net with Gaussian kernels. Two signal Gaussians, 5000 events
each, with means $(-5,-5)$ and $(5,-5)$, respectively, and identity
covariance matrices (top left). RBF net output for signal (solid
red) and background (hollow blue) events (top right). True signal
events selected by requiring the neural net output to be above 0.5
(bottom left). True background events selected by the same requirement
(bottom right). The background density, not shown here, is uniform on
the square $(-10,-10)-(10,10)$.}

\subsection{Decision Trees}
\label{sec:trees}

A decision tree recursively splits training data into rectangular
regions (nodes). It starts with all input data and looks at all
possible binary splits in each dimension to select one with a highest
figure of merit. Then the tree examines each of the obtained nodes and
splits it into finer nodes. This procedure is repeated until a
stopping criterion is satisfied.

Each binary split is optimized using a certain \fom. Suppose we are
solving a classification problem with only two categories, signal and
background. Suppose we start with a parent node with the total weight
of events given by $W$ and split it into two daughter nodes with
weights $W^{(1)}$ and $W^{(2)}=W-W^{(1)}$, respectively. The parent
node was labeled as either ``signal'' or ``background'', and we assign
temporary labels to the two daughter nodes; the daughter labels can
be, of course, swapped if this results in a better \fom. Let $p$ and
$q=1-p$ be fractions of correctly classified and misclassified events
in each node, with proper indices supplied for the daughter
nodes. Three popular figures of merit, $Q$, used in conventional
classification trees are given by
\begin{itemize}
	\item Correctly classified fraction of events: $Q(p,q)=p$.
	\item Negative Gini index: $Q(p,q)=-2pq$.
	\item Negative cross-entropy: $Q(p,q)=p\log p+q\log q$.
\end{itemize}
A split is optimized to give the largest overall \fom,
\begin{equation}
Q_{\mbox{split}} = \frac{W^{(1)}Q_1+W^{(2)}Q_2}{W}\ , 
\end{equation}
where $Q_1$ and $Q_2$ are figures of merit computed for the two
daughter nodes.  If the initial \fom\ $Q$ cannot be improved, this
node is not split and becomes ``leaf'', or ``terminal''. A terminal
node is labeled as ``signal'' if the total weight of signal events
contained in this node is not less than the total weight of background
events in this node.

Several criteria can be used to stop training; for a brief review see,
e.g., a section on decision trees in Ref.~\cite{kuncheva}. Only one
stopping criterion is implemented in \spr: the user must specify the
minimal number of events per tree node. The tree continues making new
nodes until it is composed of leaves only --- nodes that cannot be
split without a decrease in the \fom\ and nodes that cannot be split
because they have too few events.

Note that the quantities $p$ and $q$ used for the split optimization
were defined with no regard to event categories. A conventional
decision tree makes no distinction between signal and background and
spends an equal amount of time optimizing nodes dominated by signal
and background events. In HEP analysis, one most usually treats the
two categories asymetrically, i.e., one is only concerned with
optimizing signal but not background. If $w_1$ and $w_0$ are weights
of the signal and background components, respectively, in a given
node, one can use the following asymmetric optimization criteria
included in \spr:
\begin{itemize}
	\item Signal purity: $Q(w_1,w_0)=w_1/(w_1+w_0)$.
	\item Signal significance: $Q(w_1,w_0)=w_1/\sqrt{w_1+w_0}$.
	\item Tagging efficiency: 
	$Q(w_1,w_0)=(w_1+w_0)\left[ 1 - 2w_0/(w_1+w_0) \right]_+^2$.
\end{itemize}
The ``+'' subscript above indicates that the expression in the
brackets is used only when it is positive; at negative values it
yields zero. 
%A node is labeled as ``signal'' if $Q(w_1,w_0)\geq
%Q(w_0,w_1)$ and ``background'' otherwise. 
The \fom\ for split
optimization then becomes
\begin{equation}
Q_{\mbox{split}} = \max(Q_1,Q_2)\ . 
\end{equation}
Similarly, if the parent \fom\ $Q$ cannot be improved, this node is
labeled as ``terminal''.  The algorithm for class assignment to
terminal nodes will be discussed in detail later in this section.

The choice of an optimization criterion is not
straightforward. Consider, for example, separating the two bivariate
Gaussians from uniform background shown in
Fig.~\ref{fig:tree_twogauss_and_uniform}. A decision tree based on the
signal significance finds a big rectangle that covers both
Gaussians. At the same time, a tree using the Gini index puts a
separate rectangle around each Gaussian. Although the Gini-based tree
yields a slightly worse (by 4\%) signal significance, it provides more
insight into the data structure. A purity-based tree makes many small
terminal nodes with high signal-to-background ratios and is hardly
appropriate for this problem. However, if one wants to search for
small pure signal clusters, the purity criterion comes in handy, as
discussed in Section~\ref{sec:hunter}.

\simplex{htbp}{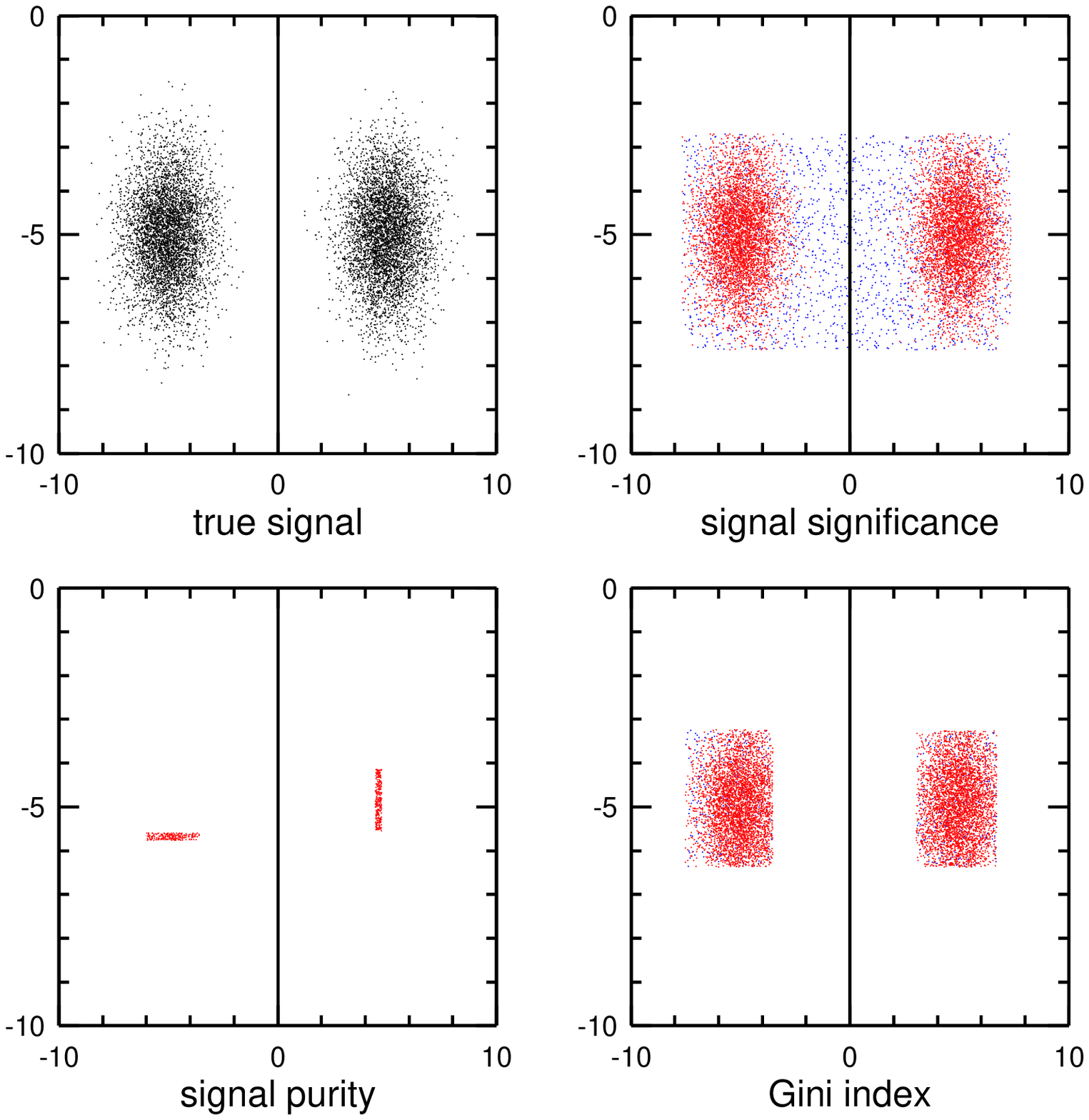}{tree_twogauss_and_uniform}{
Separation of the two signal Gaussians from uniform background by
classification trees. True signal events drawn from two bivariate
Gaussians with means $(-5,5)$ and $(5,5)$, respectively, and identity
covariance matrices (top left). The most significant terminal node of
the tree based on the signal significance (top right). Two most
significant terminal nodes of the purity-optimized tree with at least
200 events per node (bottom left). Two terminal nodes of the tree
based on the Gini index with at least 4350 events per node (bottom
right). True signal events are shown in red, and true background
events are shown in blue. The minimal node size for each tree was
chosen by maximizing the selected \fom\ on a validation set.}

Two popular commercial decision trees, CART~\cite{cart} and C4.5
developed on the basis of the ID3 algorithm~\cite{c45}, are
undoubtedly implemented at a higher level of sophistication. CART, for
example, allows to make splits on linear combinations of input
variables. To simplify the tree architecture, CART deploys a pruning
algorithm after all terminal nodes have been found. Suppose that a
non-terminal node in a tree is connected to $T$ terminal nodes with
optimized figures of merit $Q_t$ and weights $W^{(t)};\
t=1,...,T$. The cost complexity criterion for this non-terminal node
is defined as 
\begin{equation}
Q_{\mbox{cost}} = \frac{\sum_{t=1}^T W^{(t)}Q_t - \alpha T}{W}\  , 
\end{equation}
where $\alpha>0$ is the cost parameter and $W=\sum_{t=1}^T W^{(t)}$ is
the total weight of the terminal nodes. If the cost complexity
criterion is less than the \fom\ for this non-terminal node $Q$, the
node is declared ``terminal'' and all its daughters are discarded.

Decision trees are not pruned in the \spr\ implementation. It is
possible, however, to merge terminal signal nodes if requested by the
user. The merging procedure selects a subset of terminal nodes to
optimize the overall \fom. Searching through all possible combinations
of terminal nodes would be CPU-exhaustive, especially for a large
number of nodes; hence, a more efficient algorithm is deployed.
First, all signal terminal nodes obtained in the training phase are
sorted by signal purity in descending order. The algorithm then
computes the overall \fom\ for the $n$ first nodes in the sorted list
with $n$ taking consecutive values from 1 to the full length of the
list. The optimal combination of the terminal nodes is given by the
highest \fom\ computed in this manner. It can be shown that this
algorithm selects the optimal subset of the terminal nodes for any
\fom, $Q=Q(w_1,w_0)$, which increases with purity, $P=w_1/(w_1+w_0)$,
if the total event weight, $W=w_1+w_0$, is kept constant. Rigorous
proof of this statement is beyond the scope of this paper.

If the user wants to have terminal signal nodes merged, the tree
assigns all terminal nodes with a non-zero signal contribution to the
signal category. Terminal nodes that do not contribute to the overall
\fom\ are rejected by the merging procedure.

In effect, the \spr\ implementation of decision trees offers two
different algorithms. The first algorithm, close to that used by
conventional decision trees such as CART and C4.5, chooses a symmetric
\fom\ and finds signal and background terminal nodes. In this case terminal 
nodes are not merged. The second algorithm, suited specifically for
HEP analysis, uses an asymmetric \fom, assigns all terminal nodes with
non-zero signal content to the signal category and never attempts to
find regions with high background purity. Terminal nodes are then
merged to get rid of those that do not contribute to the overall signal
\fom. The distinction between these two algorithms is not enforced 
by the package. For example, the user can choose a symmetric
optimization criterion and merge terminal nodes or choose an
asymmetric criterion without merging. These possibilities, however, do
not make much sense for most practical problems. To summarize --- if
you choose a symmetric \fom\ such as the correctly classified
fraction, Gini index or cross-entropy, do not merge terminal nodes; if
you choose an asymmetric \fom\ such as the signal purity or signal
significance, make sure to merge terminal nodes.

Compared to neural nets, decision trees typically offer an inferior
predictive power. Their main strength is interpretability. Rectangular
regions can be easily understood in many dimensions. Decision trees
that allow splits on linear combinations of variables give a better
predictive power but at the same time, following the Breiman
uncertainty principle, become harder to interpret.

Another disadvantage of decision trees is the lack of training
stability. Binary splits cannot be reversed. If a tree cuts too deep
into the signal region, it will never recover and it will fail to find
a small subset of optimal nodes. This is especially true for trees without
pruning.

Due to the simplicity of the training mechanism, decision trees are
less fragile than neural nets. They can easily deal with strongly
correlated variables and mixed data types.

The three-stage training-validation-test routine is absolutely
necessary for decision trees. If the tree is constructed by maximizing
the chosen \fom\ on the training set, it will most likely be
overtrained. At the same time, the \fom\ computed for the validation
set can be substantially larger than that for the test set.

Several applications of decision trees to HEP
analysis~\cite{treesinhep} gave optimistic conclusions about their
predictive power. While it is entirely plausible that in some cases
decision trees can compete with neural nets and even exceed their
predictive power, a significantly better performance can imply that
the neural net was not properly trained. The last publication in
Ref.~\cite{treesinhep} is particularly alarming in this
sense. According to this paper, CART gives a spectacular improvement
over the neural net for the quality of the $\pi/K$ separation at the
\babar\ detector. However, the paper says close to nothing about what
neural net was used and how it was trained. It is not even clear if
the neural net and CART were trained on the same set of input
variables and similar data sets. About 70 input variables, both
discrete and continuous, were used for CART optimization in this
analysis; some of them are strongly correlated. This is a hard problem
for neural nets --- processing a large training sample with dozens of
input variables can be excruciatingly slow, and the presence of mixed
inputs and strongly correlated variables can make the training
procedure unstable. To obtain an accurate estimate of the neural net
performance, one would have to get rid of the strongly correlated
variables, exclude inputs that do not contribute much to the
predictive power, and try several network configurations. Without this
exercise, such a comparison is hardly useful.

\subsection{Bump Hunting}
\label{sec:hunter}

Another useful tool implemented in \spr\ is PRIM, a bump hunting
algorithm~\cite{prim}. PRIM searches for rectangular regions with an
optimal separation between the categories but, unlike a decision tree,
it follows a more conservative approach. Instead of recursively
splitting the input space into finer nodes, this algorithm attempts to
find one rectangular region with a globally optimized \fom. This
search is implemented in two steps:
\begin{itemize}
	\item Shrinkage. At this stage, the bump hunter gradually
	reduces the size of the signal box by imposing binary
	splits. The most optimal split is found at each iteration by
	searching through all possible splits in all input
	variables. The rate of shrinkage is controlled by a ``peel''
	parameter, the maximal fraction of events that can be peeled
	off the signal box with one binary split. This parameter is
	supplied by the user and can be used to adjust the level of
	conservatism. If the bump hunter cannot find a binary split to
	improve the \fom, shrinkage is stopped.  
	\item Expansion. At this stage, the hunter attempts to relax
	the bounds of the signal box to optimize the \fom.
\end{itemize}
After the signal box has been found, the hunter removes points located
inside this box from the original data set and starts a new search
from scratch.

The bump hunting algorithm is slower and more conservative than the
recursive splitting used by decision trees. It is most suitable for
problems where the user wants to find a pre-determined number of
signal regions. For example, if you wish to find one signal box, the
bump hunter is the right tool. A decision tree can be forced to
produce only one signal terminal node too --- by requesting a large
enough minimal size of a terminal node. But the bump hunter is more
flexible. By trying various peel parameters, one can find a box close
to the optimal one.

An interesting application of bump hunting is exploratory search for
new signatures in a multidimensional space. The Sleuth
algorithm~\cite{sleuth} developed by the D0 Collaboration searches for
new signatures by splitting data into Voronoi cells and identifying
cells with low probabilities of observing that many events given the
Monte Carlo expectation. A closely related problem is estimation of
the goodness of fit~\cite{my_gof} by comparing minimal and maximal
sizes of observed event clusters to those expected from Monte Carlo
simulation. These methods unfortunately rely on a uniformity
transformation from the space of physical observables into the
flattened space where Voronoi cells or clusters are constructed. A
uniformity transformation is not unique. The choice of this
transformation can have a serious impact on the result, especially in
multidimensional problems. With bump hunting, a uniformity
transformation is not necessary. One simply labels Monte Carlo events
as category 0 and observed data as category 1, and then searches for
bumps in the space of unaltered physical variables.

It is surprising how good are the bump hunter and decision trees for
locating even tiny clusters of events. Suppose we expect 10,000 points
drawn from a uniform 2D distribution within a unit square
$(0,0)-(1,1)$ and observe 9910 uniformly distributed points with two
tiny peaks included: a 60 event peak located at $(0.25,0.25)$ and a 30
event peak located at $(0.75,0.75)$. The standard deviations for both
peaks are 0.05 and there is no correlation between the variables. The
two bumps are hardly visible on the scatter plot shown in
Fig.~\ref{fig:tiny_bumps_true_signal}: the larger bump can be seen but
the smaller bump is practically lost among uniformly distributed
points.  We now attempt to find these tiny bumps with a decision tree
and a bump hunter. The most efficient \fom\ for this task is the
signal purity. The decision tree easily finds the larger bump but
typically misses the smaller bump at $(0.75,0.75)$. The best result is
obtained with a decision tree with at least 30 events per node; the
smaller bump in this case is found as the 6th most significant
terminal node and it is stretched vertically bearing little
resemblance to the real bump. The bump hunter is more
flexible. Depending on the peel parameter and the requested minimal
bump size, one can obtain a variety of bump configurations. For
example, the bump hunter with at least 20 events per node and peel
parameter 0.7 places the most significant bump with 20 signal and 1
background events in the vicinity of $(0.75,0.75)$ and locates several
bumps at $(0.25,0.25)$. Note that the most significant bumps are not
necessarily found in the order of their significance. The most
significant bump marked in red is only the 11th found by the hunter.

\simplex{htbp}{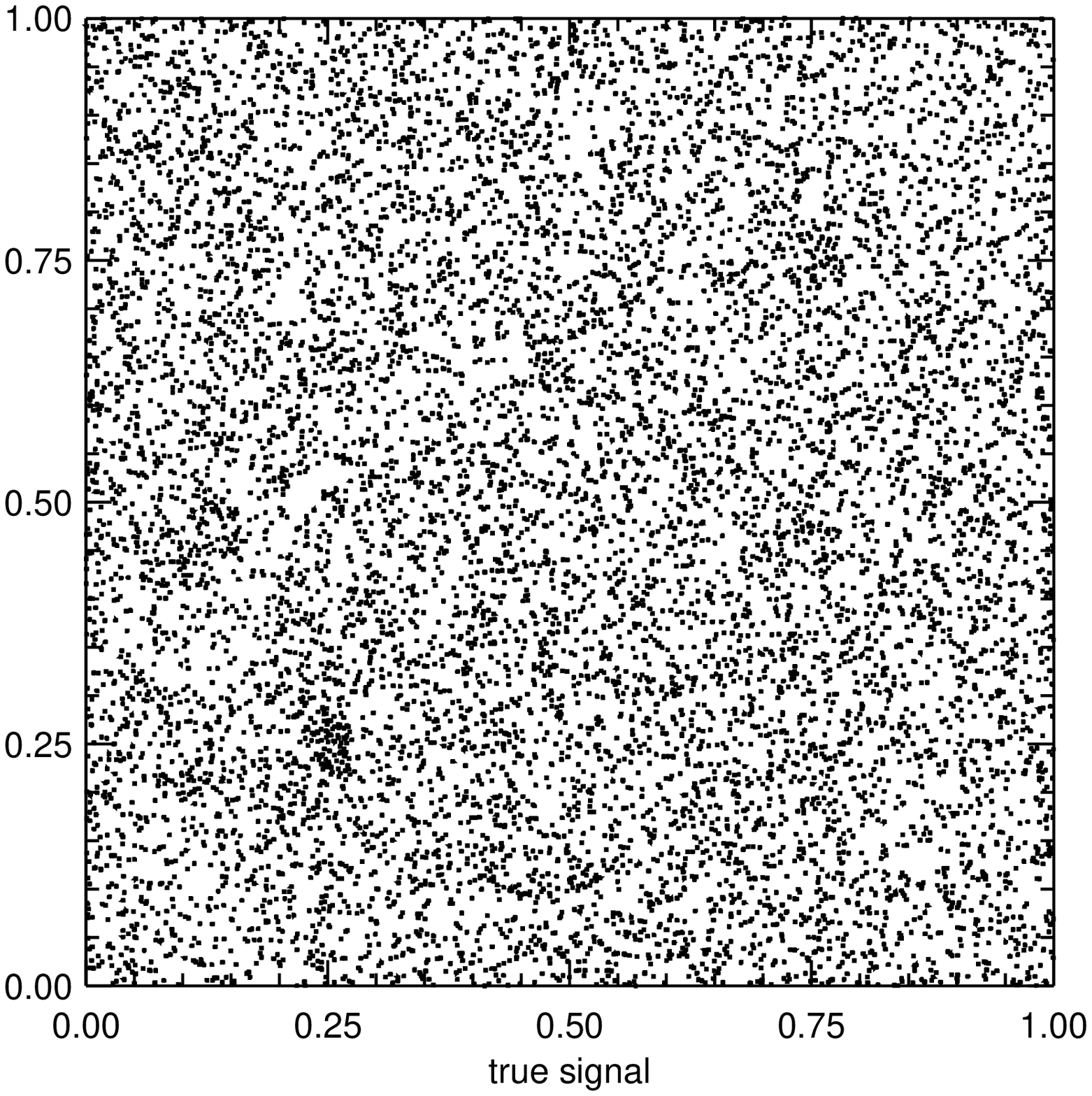}{tiny_bumps_true_signal}{
Distribution of signal events for the bump hunting problem: 9910
events uniformly distributed on the unit square, 60 events drawn from
a two-dimensional Gaussian centered at $(0.25,0.25)$ with standard
deviation 0.05 on each axis, and 30 events drawn from a
two-dimensional Gaussian centered at $(0.75,0.75)$ with standard
deviation 0.05 on each axis. Honestly, can you see the peaks?}

\duplex{htbp}{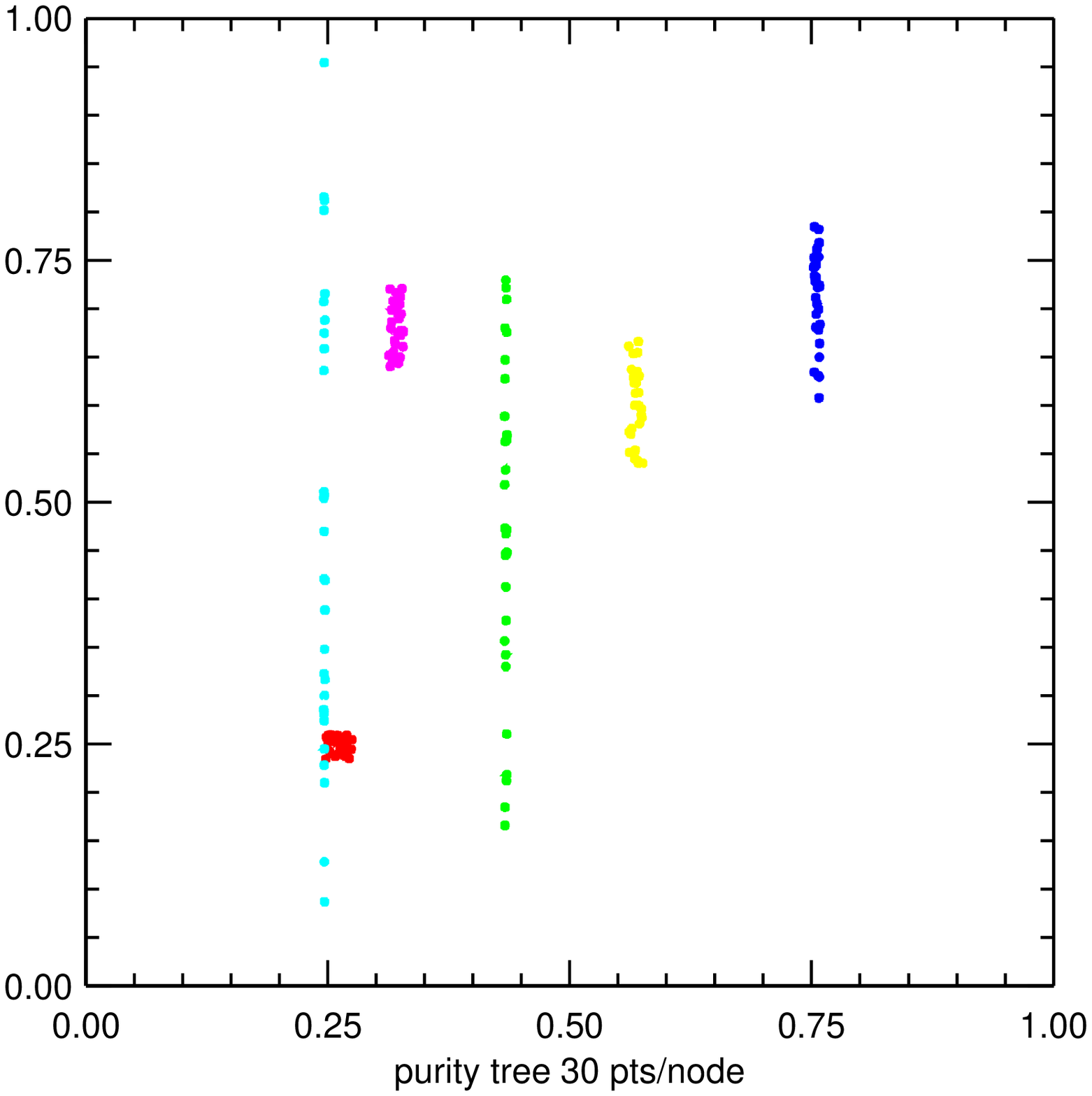}{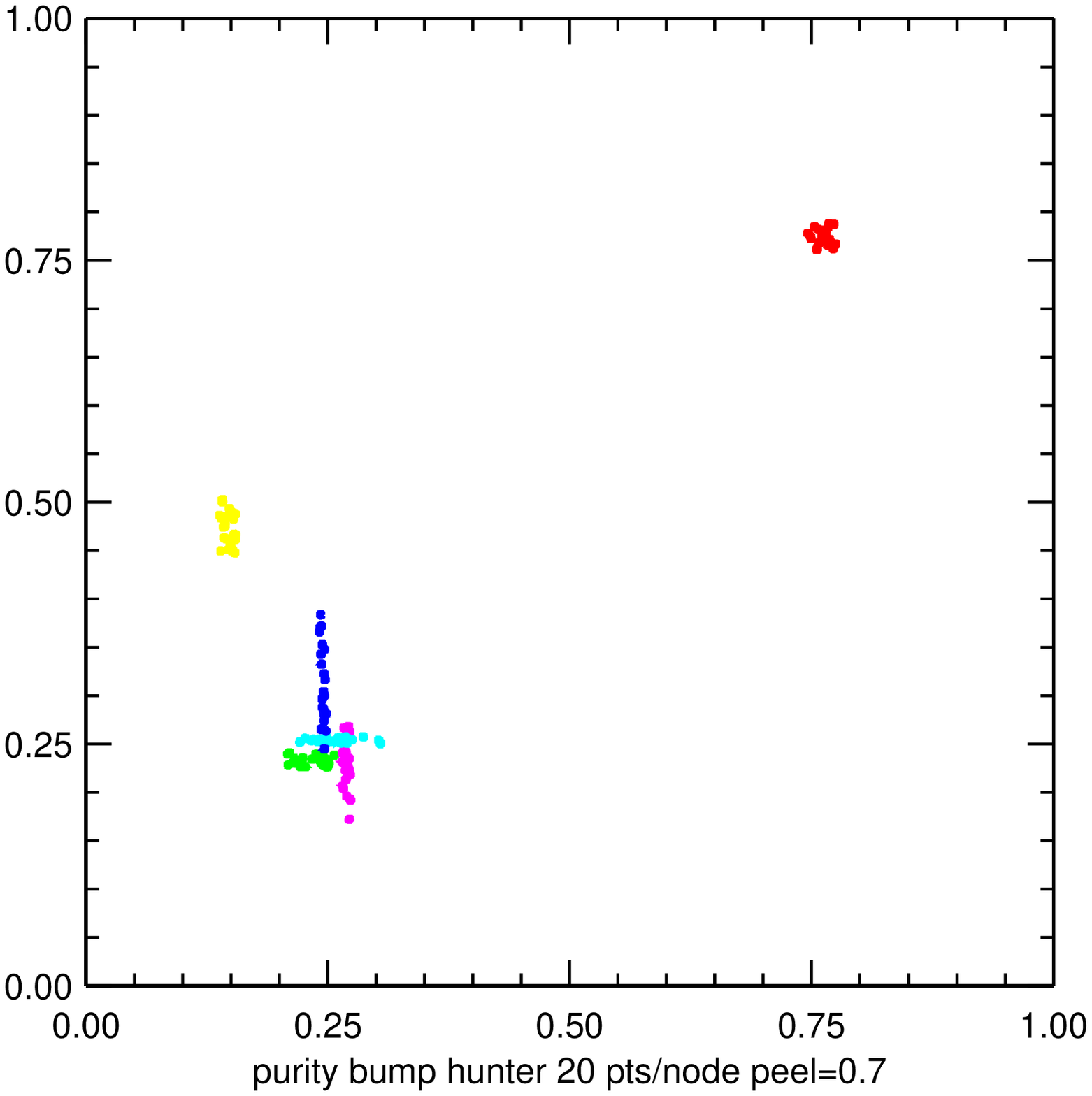}
{tiny_bumps_tree_and_hunter} {Six most significant terminal nodes
produced by a purity-based decision tree with at least 30 points per
node (left) and six most significant bumps produced by a purity-based
bump hunting algorithm with at least 20 points per node and peel
parameter 0.7 (right). The nodes/bumps are sorted by their purity in
the descending order as follows: red (most significant), yellow,
green, magenta, cyan, and blue (least significant). The bump hunting
algorithm is somewhat more efficient for finding the two signal peaks
at $(0.25,0.25)$ and $(0.75,0.75)$.}

\subsection{AdaBoost}
\label{sec:ada}

The training instability of decision trees can be alleviated by using
a more flexible training algorithm. Instead of imposing irreversible
hard splits, one can impose ``soft'' splits. The final category is
then assigned to an event using a weighted vote of all soft
splits. This classifier replaces the hard decision, signal versus
background, with a continuous output obtained by summation of weights
from relevant splits. AdaBoost is one popular implementation of this
approach.

AdaBoost works by enhancing weights of misclassified events and
training new splits on the reweighted data sample. At the first step,
all weights are set to initial event weights in the training sample of
size $N$:
\begin{equation}
w_n^{(0)} = w_n;\ \ \ n=1,...,N.
\end{equation}
At iteration $k$, the algorithm imposes a new binary split $\eta_k(x)$
to minimize the misclassified fraction of events and computes the
associated misclassification error:
\begin{equation}
\epsilon_k = \frac{\sum_{n=1}^N w_n^{(k-1)} I(\eta_k(x_n)\neq y_n)}
             {\sum_{n=1}^N w_n^{(k-1)}}\ ,
\end{equation}
where $I$ is an indicator function equal to 1 if the indicator
expression is true and 0 otherwise, $y_i$ is the true category of
event $n$ (-1 for background and 1 for signal), and the binary split
$\eta_k(x)$ is a function with discrete output, equal to 1 if an event
is accepted by the split and -1 otherwise. The weights of misclassified
events are then enhanced:
\begin{equation}
w_n^{(k)} = \frac{w_n^{(k-1)}}{2\epsilon_k}\ ,
\end{equation}
and the weights of correctly classified events are suppressed:
\begin{equation}
w_n^{(k)} = \frac{w_n^{(k-1)}}{2(1-\epsilon_k)}\ .
\end{equation}
To impose splits, the algorithm loops through data dimensions
cyclically --- if the $k$th split was imposed on dimension $d$, the
$(k+1)$st split will be imposed on dimension $d+1$, and after the last
dimension has been processed, the algorithm goes back to the first
dimension. This iterative process is continued as long as the
misclassification error at iteration $k$ is less than 50\%,
$\epsilon_k<0.5$, and the number of training splits requested by the
user is not exceeded. After the training has been completed, a new
event is classified by a weighted vote of all binary splits:
\begin{equation}
\label{eq:ada}
f(x) = \sum_{k=1}^K \beta_k \eta_k(x)
\end{equation}
with the beta weight of each split given by
\begin{equation}
\beta_k = \log \left( \frac{1-\epsilon_k}{\epsilon_k} \right)\ .
\end{equation}
For a large number of binary splits $K$, the function $f(x)$ is
practically continuous. It gives negative values for background-like
events and positive values for signal-like events.

Formally, the AdaBoost algorithm described above can be derived by
minimization of the exponential loss
\begin{equation}
{\mathcal E} = \sum_{n=1}^N \exp( -y_n f(x_n) )\ .
\end{equation}
Phenomenologically, AdaBoost can be understood as a clever mechanism
which improves the quality of classification by enhancing poorly
classified events and training new splits on these events. The trained
splits are then rated by their quality: if the misclassification error
for split $k$ is small, this split is included with a large weight
$\beta_k$; if the misclassification error is barely better than 50\%,
this split will only have a small effect on the output.

Note that above I used a different convention for signal and
background labeling --- unlike in the rest of the paper, background is
labeled as -1, not 0. This was done simply to preserve the notation in
which AdaBoost is explained in statistical textbooks and to illustrate
the concept of the exponential loss. To maintain consistency with
other methods, the AdaBoost implementation in \spr\ uses the
conventional labeling (0 for background) and normalizes the beta
weights to a unit sum. The AdaBoost output from Eqn.~(\ref{eq:ada}) is
therefore shifted by 0.5 and confined between 0 and 1. Signal-like
events populate the region above 0.5 and background-like events
populate the region below 0.5.

AdaBoost has been discussed here as a method of consecutive
application of binary splits. The AdaBoost algorithm, of course, can
be as well used with any other classifiers. \spr\ includes AdaBoost
implementations with binary splits, with linear and quadratic Fisher
discriminants, and with decision trees.

AdaBoost combines the best features of decision trees and neural nets
--- needless to say, at the expense of interpretability. The training
algorithm is very robust. AdaBoost can easily deal with mixed discrete
and continuous input variables, and it gives a high predictive
power. Unlike neural nets, AdaBoost does not require any pre-training
initialization. AdaBoost generally takes less training time than the
two neural nets discussed in this note. This advantage in the training
speed becomes spectacular in high-dimensional problems. For example,
the training time for AdaBoost with soft binary splits scales linearly
with the number of dimensions, while for the standard backpropagation
neural net this dependence is at best quadratic and can be much worse
if more than one hidden layer is used.

Separation of two bivariate Gaussians from uniform background by
AdaBoost with 500 soft binary splits on each input variable is shown
in Fig.~\ref{fig:ada_twogauss_and_uniform}.

\simplex{htbp}{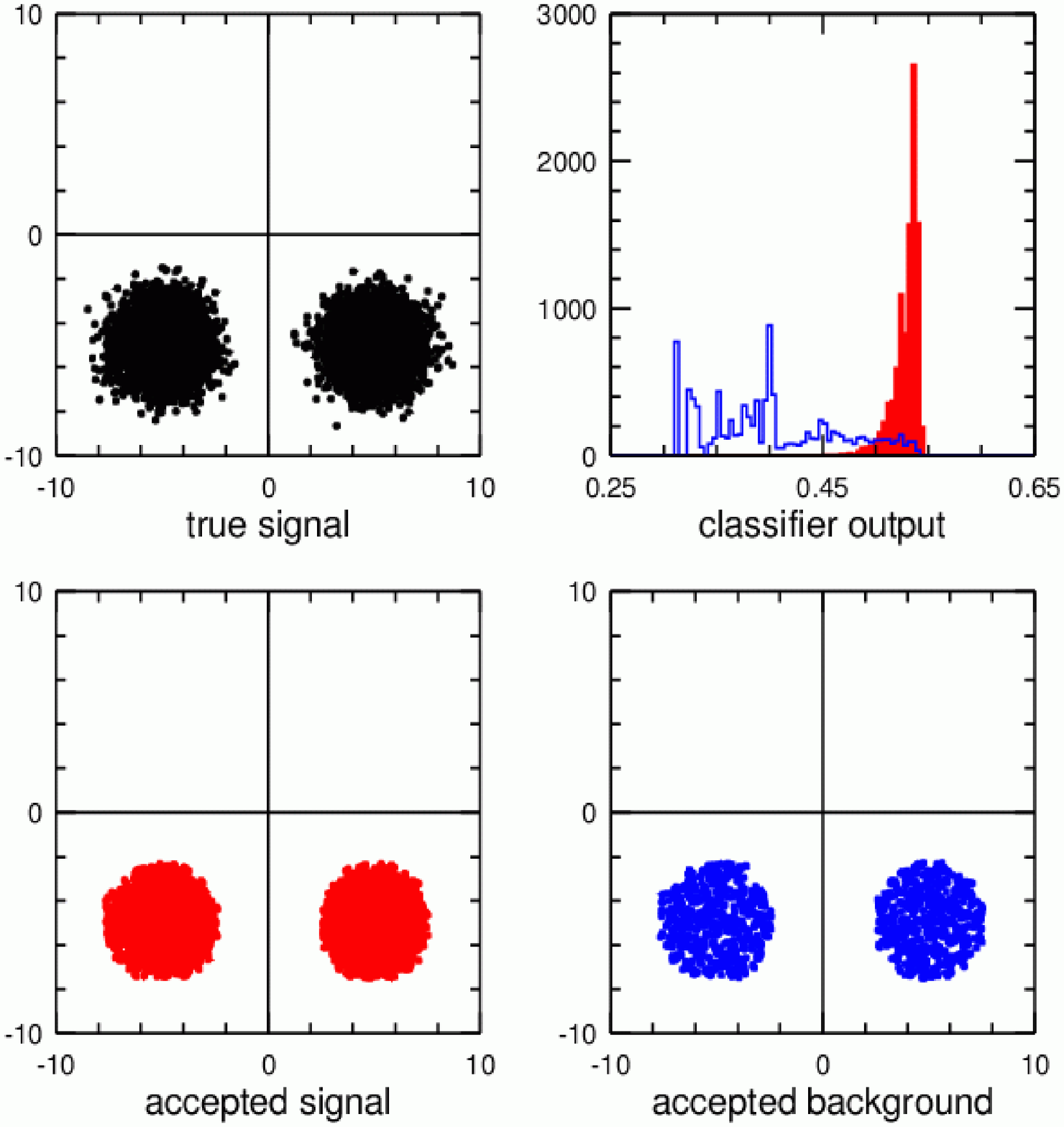}{ada_twogauss_and_uniform}{
Separation of two bivariate Gaussians from uniform background by
AdaBoost with 500 soft binary splits on each input variable. Two
signal Gaussians, 5000 events each, with means $(-5,-5)$ and $(5,-5)$,
respectively, and identity covariance matrices (top left). AdaBoost
output for signal (solid red) and background (hollow blue) events (top
right). True signal events selected by requiring the AdaBoost output
to be above 0.5 (bottom left). True background events selected by the
same requirement (bottom right). The background density, not shown
here, is uniform on the square $(-10,-10)-(10,10)$.}

\subsection{Bagging and Random Forest}
\label{sec:bagging}

Another powerful tool for combining weak classifiers is bootstrap
aggregating, or ``bagging''~\cite{bagging}. Also implemented in \spr,
this method will be described in more detail in a separate note
submitted simultaneously.

\subsection{Combining Classifiers}
\label{sec:combiner}

Classifiers independently trained on data subsets can be combined to
obtain a higher overall predictive power. Methods for combining
classifiers have been a subject of recent
research~\cite{kuncheva}. One approach is to train one global
classifier in the space of subclassifier outputs. For example, to
discriminate signal from background, one could look at various sources
of background and train a separate classifier for each data subset
composed of the signal and a corresponding background source. Then
output values of all individual classifiers computed for the full
training set that includes all sources of background could be
processed by one global classifier. \spr\ implements a classifier
combiner using AdaBoost. The user supplies trained subclassifiers to
the combiner, and the combiner runs the AdaBoost algorithm, either
with soft binary splits or with Fisher discriminants or with decision
trees, using output values of the subclassifiers as input data.

I would like to stress the distinction between AdaBoost and the
classifier combiner described in the previous paragraph. The former
combines a large number of weak classifiers by applying them
sequentially. The latter combines a few powerful classifiers by
training another powerful classifier in the space of their output
values.

\section{Other Methods}
\label{sec:other}

A few other methods of general use for statistics analysis in HEP have
been implemented in \spr. 

Computation of data moments is sometimes necessary in analysis
applications. Mean, variance and kurtosis for variables and their
combinations can be computed. A test of independence of two variables
drawn from a joint elliptical distribution\footnote{A distribution is
called ``elliptical'' if its density is constant on the surface
$x^TAx$ with a positive definite matrix $A$. This class obviously
includes multivariate Gaussian distributions.} can be performed using
a simple analytic formula~\cite{anderson}.

Bootstrap~\cite{bootstrap} is a convenient tool that can be used to
estimate the distribution of a statistic computed for a small data set
if Monte Carlo or other means of sample regeneration are not
available. This can be done by sampling from the data set with
replacement. To build one bootstrap replica, one needs to draw $N$
events with replacement out of the data set of size $N$. To study the
distribution of the statistic of interest, one typically needs to
build 100-200 bootstrap replicas. This method can be efficiently used
for data sets of size $N=20$ or larger.

\begin{figure}[htbp]
   \begin{center}
%   \vskip -2.0 cm
   \hbox{ \quad \parbox[t]{14.5cm}{
   \psfig{file=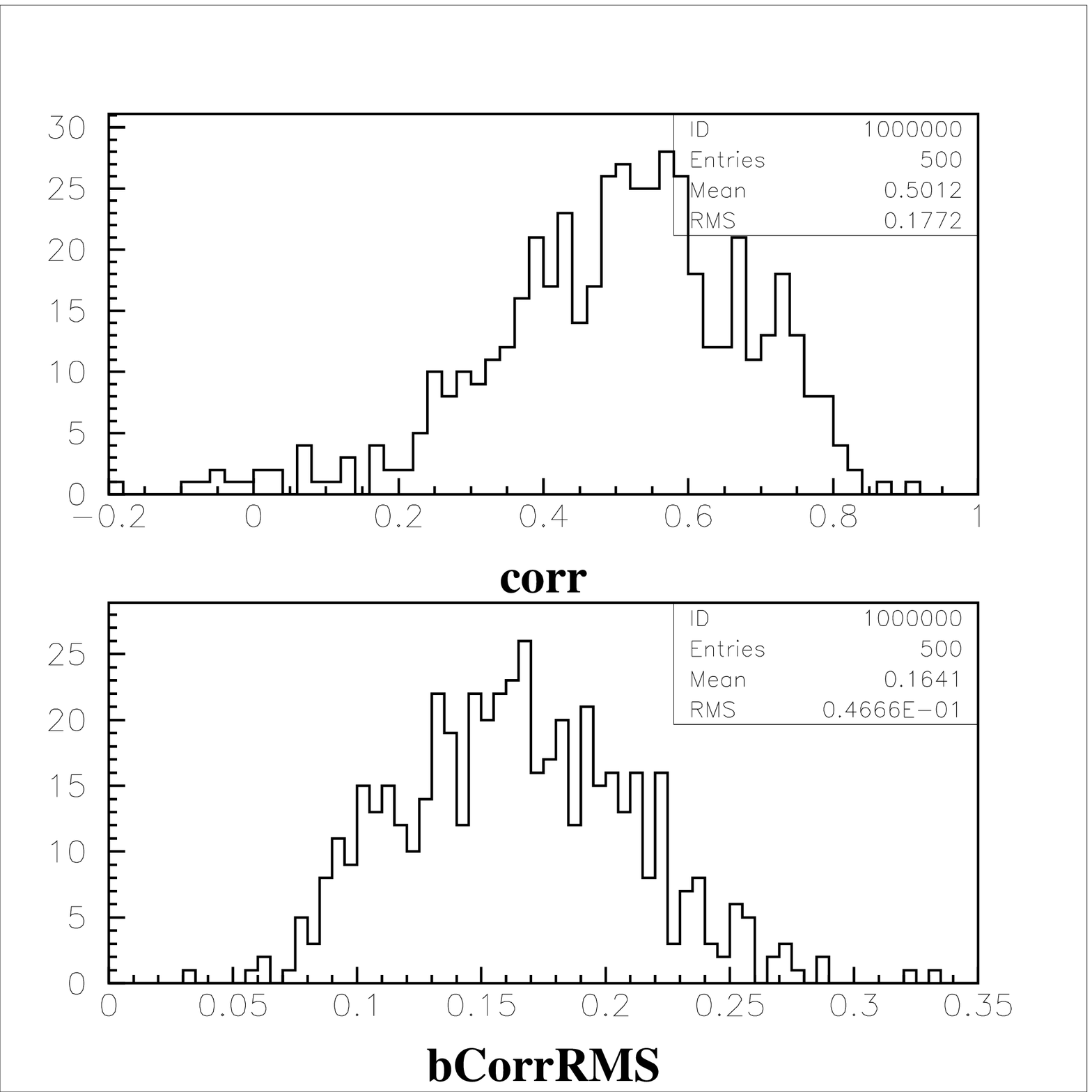,width=6.5cm} \caption[c]{\small
   \label{fig:bootstrap} Bootstrap analysis of 500 estimators of the
   correlation coefficient computed for samples of 20 events drawn
   from a bivariate Gaussian distribution described in the
   text. Distribution of 500 correlation estimates
   (above). Distribution of 500 bootstrap estimates of the correlation
   estimator standard deviation obtained by generating 100 bootstrap
   replicas for each sample of 20 events (bottom). The r.m.s. of the
   top distribution is fairly close to the mean of the bottom
   distribution. This shows that the bootstrap method on average gives
   a decent estimate of the standard deviation of the correlation
   coefficient estimator.  } } } \quad \end{center}
\end{figure}

Fig.~\ref{fig:bootstrap} illustrates bootstrap application for
computation of the error of a correlation coefficient estimator. Two
random variables, $x$ and $y$, have a joint bivariate Gaussian
distribution with a covariance matrix $\Sigma$:
$\Sigma_{11}=\Sigma_{22}=1$; $\Sigma_{12}=\Sigma_{21}=0.5$. The
correlation coefficient between the two variables is estimated for a
sample of 20 events in the usual way: 
$\hat{\rho} = \left( \sum_{i=1}^{20} (x_i-\bar{x})(y_i-\bar{y}) \right) / 
\left( \sqrt{\sum_{i=1}^{20} (x_i-\bar{x})^2} 
\cdot \sqrt{\sum_{i=1}^{20} (y_i-\bar{y})^2} \right)$, where
$\bar{x}$ and $\bar{y}$ are the two means. Imagine that you only have
a data set with 20 events and you do not know the underlying
distribution. How would compute the error associated with the
estimator $\hat{\rho}$? The usual approach in HEP analysis is to
assume an underlying distribution for $x$ and $y$ and study the
distribution of $\hat{\rho}$ based on this assumption. Bootstrap
offers a non-parametric and assumption-free approach --- the
distribution of $\hat{\rho}$ is studied on bootstrap replicas. As
shown in Fig.~\ref{fig:bootstrap}, bootstrap provides an unbiased
estimate of the error of $\hat{\rho}$ based on 100 replicas for
each data set.

\section{A C++ Implementation: Object-Oriented Design}
\label{sec:cpp}

The methods described above have been implemented in C++ using
flexible object-oriented design patterns. A detailed description of
this implementation would be lengthy and is already included in the
{\tt README} file distributed with the package. Flexible tools have
been provided for imposing selection requirements on input data ---
the user can choose input variables, restrict analysis to a few event
categories, select events within a specified index range, and impose
orthogonal cuts on input variables. The user can easily add new
classifiers and new figures of merit for classifier optimization by
providing new implementations to the abstract interfaces set up in the
package. Executables using the new classifiers and new figures of
merit can be copied from existing executables with few modifications.

At the moment, the package accepts input data only in ascii, in a
format very similar to that used by SNNS. An optional extension to the
SNNS format is weighted input data.\footnote{SNNS assumes uniform
weights for all input events.}

External dependencies of the package include: CLHEP
libraries~\cite{clhep} used for matrix operations, CERN
libraries~\cite{cern} used for random number generation and various
probability calculations, and an internal \babar\ interface for HBOOK
or ROOT ntuple generation. If used by researchers outside \babar, the
latter will need to be replaced with a similar interface. Essentially,
only two methods need to be implemented --- one for booking an ntuple
and another one for filling ntuple columns. Because every
collaboration has a set of tools for this purpose, this should hardly
pose a problem.

\section{Separation of Signal and Background in a Search for the Radiative 
Leptonic Decay $B\to\gamma l\nu$ at \babar}
\label{sec:lnugamma}

A search for the radiative leptonic decay $B\to\gamma l\nu$ is
currently in progress at \babar; results of this analysis will be made
available to the public in the near future. This analysis focuses on
measuring the $B$ meson decay constant, $f_B$, which has not been
previously measured. While purely leptonic decays such as $B\to\mu\nu$
and $B\to e\nu$ offer the cleanest way of measuring this decay
constant, these decays are not accessible at the present level of
experimental sensitivity.  The decay $B\to\tau\nu$, with a branching
fraction of $6.6\times 10^{-5}$ predicted by the Standard Model,
suffers from substantial background due to the presence of two or
three neutrinos in the final state. Due to helicity suppression, the
decays $B\to\mu\nu$ and $B\to e\nu$ are suppressed by factors of 225
and $10^7$, respectively, and are therefore not observable at current
luminosities. The presence of the photon removes the helicity
suppression but introduces theoretical uncertainties into the
calculation of the branching fraction. Theoretical estimates of the
radiative leptonic branching fraction vary from $1.0\times 10^{-6}$ to
$4.0\times 10^{-6}$. Present upper limits on the branching fractions,
$2.0\times 10^{-4}$ and $5.2\times 10^{-5}$ for the electron and muon
channels, respectively, were set by CLEO in 1997 using 2.5~fb$^{-1}$
of on-resonance data.

Large samples of simulated Monte Carlo events are used to study signal
and background signatures in this analysis. To model the signal, about
1.2M $B\to\gamma l\nu$ signal events were generated in each
channnel. Large samples of generic $B^+B^-$, $B^0\bar{B}^0$,
$c\bar{c}$, $uds$ and $\tau^+\tau^-$ Monte Carlo events were used to
estimate background contributions. Because semileptonic decays of $B$
are an important source of background, several exclusive semileptonic
modes were simulated by Monte Carlo as well with a typical sample size
of several hundred thousand events. These modes include: $B^0\to\pi
l\nu$, $B^0\to\rho l\nu$, $B^+\to\eta l\nu$, $B^+\to\eta' l\nu$,
$B^+\to\omega l\nu$, $B^+\to\pi^0 l\nu$, and $B^+\to\rho^0 l\nu$.  For
the $B\to\gamma\mu\nu$ analysis, 224M radiative muon Monte Carlo
events were also included. Overall, 12 (13) background sources were
studied for the $B\to\gamma e\nu$ ($B\to\gamma\mu\nu$) analysis.

\begin{figure}[htbp]
   \begin{center}
   \vskip -2.0 cm
   \hbox{
   \quad
   \parbox[t]{16.5cm}{ \psfig{file=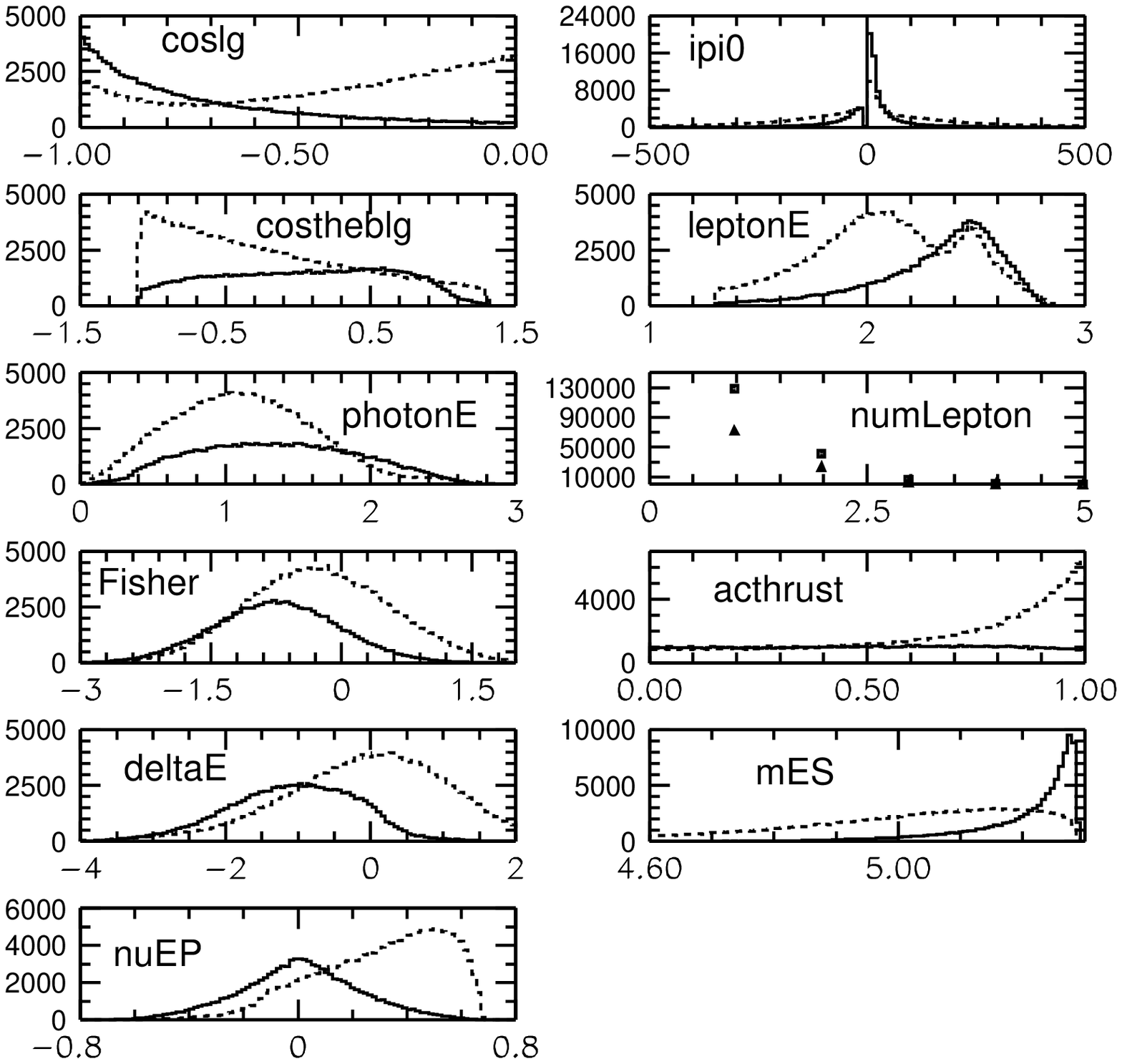,width=16.5cm}
   \caption[c]{\small  \label{fig:LNuGamma_e_Vars}  
	Separation variables for the $B\to\gamma l\nu$ analysis. 
	Signal MC is shown with a solid line 
	(triangles in the {\tt numLepton} plot), 
	and the overall combined background is shown with a dashed line
	(squares in the {\tt numLepton} plot).
	} }
   }
   \quad
   \end{center}
\end{figure}

Various preliminary requirements have been imposed to enhance the
signal purity and at the same time reduce the Monte Carlo samples to a
manageable size. These requirements include tight PID quality cuts on
the signal lepton and photon candidates and relaxed PID cuts on the
rest of the charged tracks detected by the tracking system and neutral
clusters detected by the calorimeter, a cut on the ratio of the 2nd
and 0th Wolfram moments, cuts on momenta and polar angles of the
photon and lepton candidates, a cut on the angle between the recoil
$B$ candidate and the $l\gamma$ momentum in the center-of-mass frame,
a cut on the lateral profile of the calorimeter shower used for the
signal photon reconstruction, and a fiducial cut on the direction of
the missing momentum. After these preliminary requirements have been
imposed, eleven variables are identified to be included in the final
optimization procedure: cosine between the signal lepton and photon
candidates in the center-of-mass frame ({\tt coslg}), the inverse of
the difference between the mass of the $\pi^0$ combination obtained
from the signal photon candidate and any other photon in the event
closest to the nominal $\pi^0$ mass and the nominal $\pi^0$ mass ({\tt
ipi0}), cosine of the angle between the recoil $B$ candidate and the
$l\gamma$ momentum in the center-of-mass frame ({\tt costheblg}),
energies of the lepton and photon candidates in the center-of-mass
frame ({\tt leptonE} and {\tt photonE}), the total number of
identified leptons in the event ({\tt numLepton}), a Fisher
discriminant composed of the 0th and 2nd Legendre moments of all
tracks forming the recoil $B$ candidate computed with respect to the
thrust axis given by the $l\gamma$ system ({\tt Fisher}), magnitude of
the cosine between the thrust axis of the recoil $B$ and the thrust
axis of the $l\gamma$ system ({\tt acthrust}), the difference between
the energy of the recoil $B$ candidate and the beam energy in the
center-of-mass frame ({\tt deltaE}), the beam-energy-constrained mass
of the recoil $B$ candidate ({\tt mES}), and the difference between
the missing energy and the magnitude of the missing momentum ({\tt
nuEP}). Distributions of these variables for signal and background are
shown in Fig.~\ref{fig:LNuGamma_e_Vars}. The correlation between {\tt
Fisher} and {\tt acthrust} is 0.80 for both signal and combined
background from all listed sources. {\tt costheblg} and {\tt nuEP}
show a strong negative correlation, -0.87, for the combined background
and a weaker correlation, -0.62, for the signal. All other included
variables show less significant levels of correlation.

Distributions of these eleven variables in the signal and combined
background Monte Carlo samples are used by various optimization
algorithms to maximize the signal significance expected in
210~fb$^{-1}$ of data. The training samples used for this optimization
consist of roughly half a million signal and background Monte Carlo
events in both electron and muon channels, appropriately weighted
according to the integrated luminosity observed in the data. The
training:validation:test ratio for the sample sizes is 2:1:1. The
exact breakdown of generated events for each channel is shown in
Table~I. Signal Monte Carlo samples are weighted assuming a branching
fraction of $3\times 10^{-6}$ for each channel.

Authors of this analysis deploy an original cut optimization routine
for separation of signal and background. Four variables ({\tt coslg,
numLepton, acthrust}, and {\tt mES}) are used for one-sided
optimization, and the other seven variables are used for two-sided
optimization. The available range for each variable is divided into
intervals of preselected length. At each iteration step of the
optimization procedure, a 1D plot of the signal significance versus
the cut position is constructed for each one-sided optimization
variable; for each two-sided optimization variable, a similar 2D plot
is made of the signal significance versus all possible combinations of
the two cut positions on both ends of the variable range. At each
iteration step, the optimization algorithm finds the best variable and
the best one- or two-sided cut on this variable that maximizes the
signal significance. This optimal cut on one variable is then imposed
and remains in effect for all next iterations of the algorithm. This
cut is in effect for the optimization of all selection variables
except the variable on which it has been imposed. For example, events
that are rejected by a cut on {\tt numLepton} are included in the
optimization of a new cut on {\tt numLepton} at the next iteration of
the algorithm. This allows the algorithm to relax the cut and even
remove it completely if it improves the signal significance. This
algorithm makes 10-20 iterations and produces an optimized rectangular
region in the 11-dimensional space. The optimized signal region is
then tested on independently generated test data. Note that this
algorithm does not include a validation stage.

\begin{table}[htbp]
\begin{center}
\label{tab:lnugamma_training}
\caption{Generated signal and background contributions used for the training 
samples in the $B\to\gamma l\nu$ analysis.}
\begin{tabular}{|c|c|c|}\hline
Decay               & Signal, events & Background, events \\ \hline
$B\to\gamma e\nu$   & 197k           & 352k               \\ \hline
$B\to\gamma \mu\nu$ & 161k           & 431k               \\ \hline
\end{tabular}
\end{center}
\end{table}

\begin{figure}[htbp]
   \begin{center}
   \hbox{\hskip -0cm
   \quad 
   \parbox[t]{5.5cm}{ 
     \psfig{figure=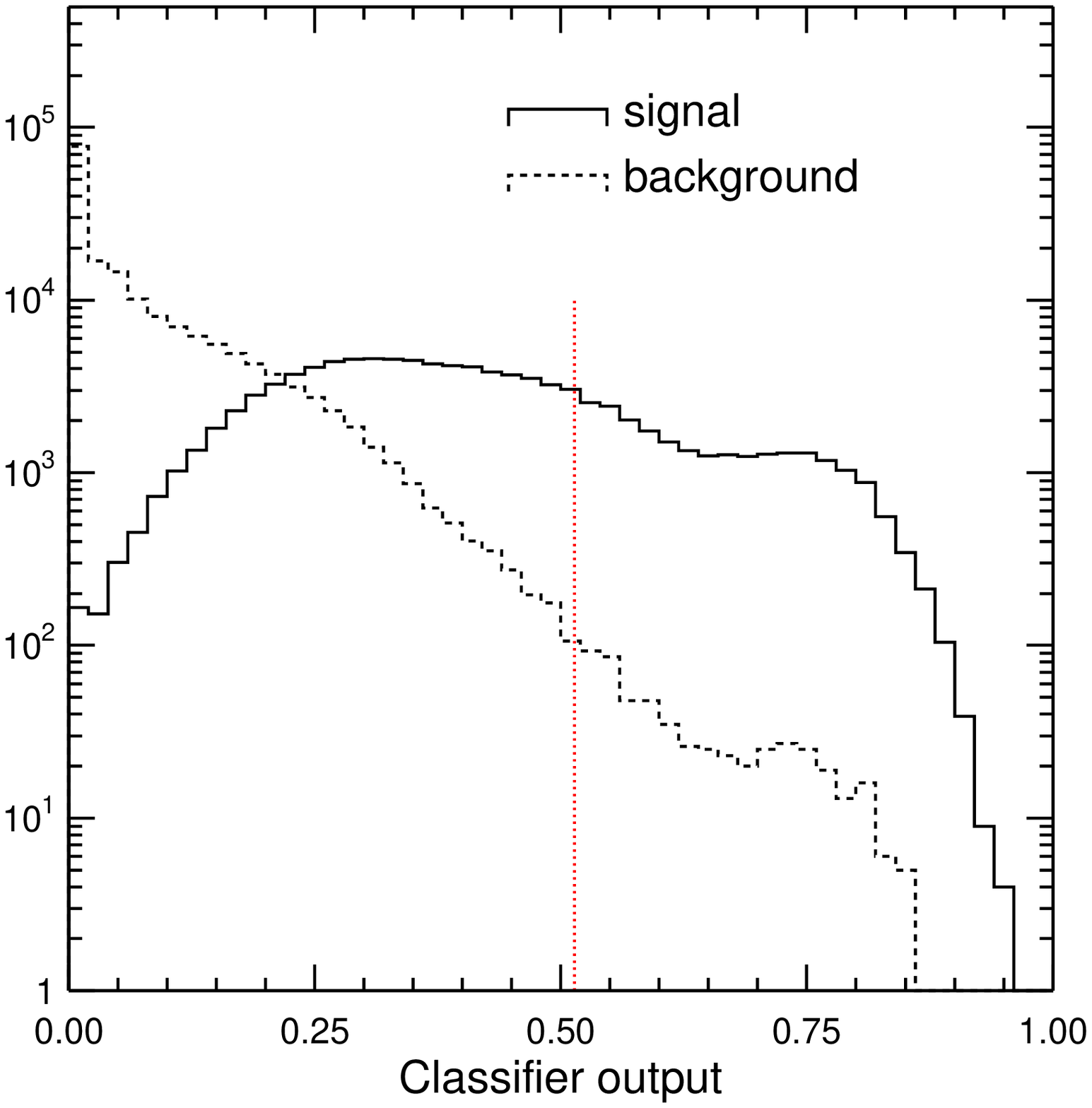,width=8cm}
   } 
   \quad
   \parbox[t]{5.5cm}{ 
     \psfig{figure=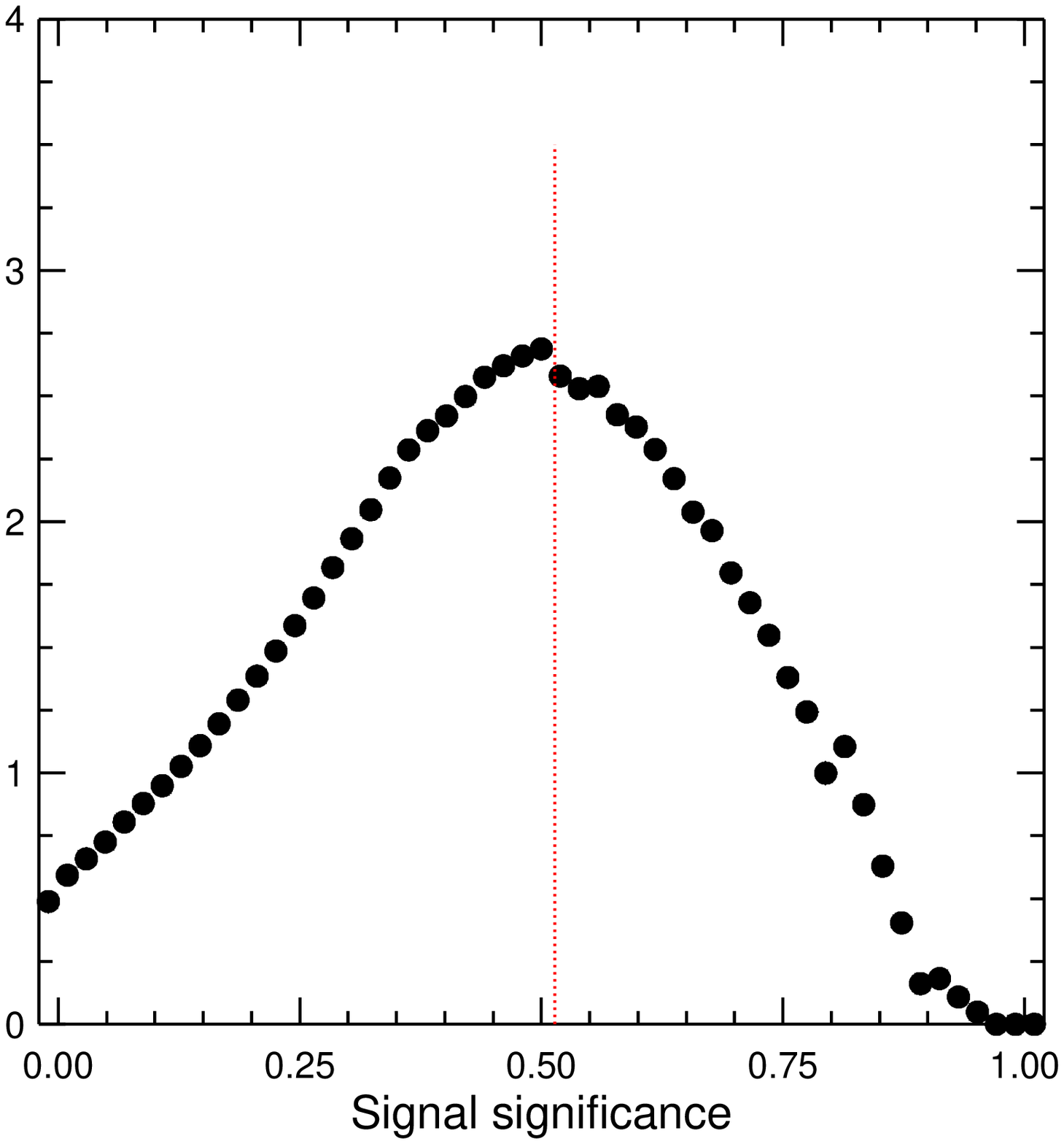,width=8cm} }
   }
   \caption[c]{\small \label{fig:LNuGamma_e_adatree} 
	Output of the AdaBoost trained 
	with 100 decision trees (left) 
	and the signal significance
	versus the lower cut on the output (right)
	for the $B\to\gamma e\nu$ test sample.
	The cut maximizing the signal significance,
	obtained using the validation sample, 
	is shown with a vertical line.
	}
   \end{center} 
\end{figure}

Besides the original method designed by the analysts, I use several
classifiers described in this note:
\begin{itemize}
	\item Decision tree optimizing the signal significance
	$S/\sqrt{S+B}$, with merged terminal nodes. The minimal node
	sizes were chosen by maximizing the signal significance on the
	validation samples to be 5k events and 13k events for the
	$B\to\gamma e\nu$ and $B\to\gamma \mu\nu$ decays,
	respectively. For each decay, the optimal tree contains two
	terminal signal nodes.

	\item Bump hunter optimizing the signal significance, with
	peel parameter 0.05, requested to find one bump only. The
	optimal value of the peel parameter was found by maximizing the
	signal significance on the validation samples.

	\item AdaBoost with binary splits. The number of binary splits
	was somewhat arbitrarily set to 700. The validation samples
	were used to find the optimal cuts on the AdaBoost output by
	maximizing the signal significance.

	\item AdaBoost with decision trees. I used 50 decision trees
	trained by optimizing the Gini index and combined them using
	the AdaBoost algorithm. Terminal nodes of the decision trees
	were not merged. The minimal node size, 100 events or,
	roughly, 0.02\% of the training sample, was found by
	maximizing the signal significance on the validation
	samples. The number of terminal signal nodes in a trained tree
	varied from 700 to 1200, except for the tree built at the very
	first iteration of the boosting algorithm, which was much
	smaller (150 signal nodes for $B\to\gamma e\nu$ and 90 signal
	nodes for $B\to\gamma \mu\nu$).

	\item AdaBoost-based combiner of subclassifiers trained on
	individual background components. I trained an individual
	AdaBoost classifier built with binary splits for each of the
	12 (13) background components in the $B\to\gamma e\nu$
	($B\to\gamma \mu\nu$) analysis. These subclassifiers were then
	combined by training a global AdaBoost with binary splits in
	the 12-(13-)dimensional space of the subclassifier
	outputs. The optimal cuts on the output of the global AdaBoost
	were found using the validation samples.
\end{itemize}
I also attempted to train a feedforward backpropagation neural net
with one hidden layer, but the network was unstable and it failed to
converge to an optimum.  The signal significance obtained for the
training, validation and test samples in both channels is shown in
Table~II. The output of the AdaBoost with decision trees, the most
powerful classifier among those included, is shown in
Fig.~\ref{fig:LNuGamma_e_adatree}.

\begin{table}[bthp]
\begin{center}
\label{tab:lnugamma_signif}
\caption{Signal significance, ${\mathcal S}_{\mbox{train}}$,
${\mathcal S}_{\mbox{valid}}$, and ${\mathcal S}_{\mbox{test}}$, for
the $B\to\gamma l\nu$ training, validation, and test samples obtained
with various classification methods. The signal significance computed
for the test sample should be used to judge the predictive power of
the included classifiers.  A branching fraction of $3\times 10^{-6}$
was assumed for both $B\to\gamma \mu\nu$ and $B\to\gamma e\nu$
decays. $W_1$ and $W_0$ represent the signal and background,
respectively, expected in the signal region after the classification
criteria have been applied; these two numbers have been estimated using the
test samples. All numbers have been normalized to the integrated
luminosity of $210\ \mbox{fb}^{-1}$. The best value of the expected
signal significance is shown in boldface.}
\begin{tabular}{|c|c|c|c|c|c|c|c|c|c|c|}\hline
Method 
& \multicolumn{5}{c|}{$B\to\gamma e\nu$} 
& \multicolumn{5}{c|}{$B\to\gamma \mu\nu$} \\ \cline{2-11}
& ${\mathcal S}_{\mbox{train}}$
& ${\mathcal S}_{\mbox{valid}}$ & ${\mathcal S}_{\mbox{test}}$ & $W_1$ & $W_0$
& ${\mathcal S}_{\mbox{train}}$
& ${\mathcal S}_{\mbox{valid}}$ & ${\mathcal S}_{\mbox{test}}$ & $W_1$ & $W_0$
\\ \hline

Original method 
& 2.66 & - & 2.42 & 37.5 & 202.2 
& 1.75 & - & 1.62 & 25.8 & 227.4 \\ \hline

Decision tree
& 3.28 & 2.72 & 2.16 & 20.3 & 68.1 
& 1.74 & 1.63 & 1.54 & 29.0 & 325.9 \\ \hline

Bump hunter with one bump 
& 2.72 & 2.54 & 2.31 & 47.5 & 376.6
& 1.76 & 1.54 & 1.54 & 31.7 & 393.8 \\ \hline

AdaBoost with binary splits 
& 2.53 & 2.65 & 2.25 & 76.4 & 1077.3 
& 1.66 & 1.71 & 1.44 & 45.2 & 935.6 \\ \hline

AdaBoost with decision trees
& 13.63 & 2.99 & {\bf 2.62} & 58.0 & 432.8
& 11.87 & 1.97 & {\bf 1.75} & 41.6 & 523.0 \\ \hline

Combiner of background subclassifiers
& 3.03 & 2.88 & 2.49 & 83.2 & 1037.2 
& 1.84 & 1.90 & 1.66 & 55.2 & 1057.1 \\ \hline

\end{tabular}
\end{center}
\end{table}

I draw several conclusions from this exercise: 
\begin{itemize}
	\item The signal region does not have a well-defined
	optimum. Using various methods, it is possible to find signal
	regions that are very different yet comparable in terms of the
	expected signal significance.

	\item From the point of view of classification, the physics
	data in this analysis are quite simple. This is why
	simple-minded classifiers such as the bump hunter and the
	original method developed by the analysts are competitive with
	flexible classifiers such as decision trees and AdaBoost.

	\item The two classifiers that require more training time,
	AdaBoost with decision trees and AdaBoost-based combiner of
	individual background subclassifiers, provide a somewhat
	better performance.

	\item AdaBoost with binary splits shows similar output
	distributions for the training and validation samples. It is
	so robust that the validation stage can be safely omitted. For
	AdaBoost with decision trees, the situation is dramatically
	different. Now the use of the validation samples is crucial
	for finding the optimal tree node size and, even more
	importantly, the optimal cut on the AdaBoost output. This
	shows the distinction between using primitive classifiers
	(binary splits) and flexible classifiers (decision trees) as
	building blocks for AdaBoost.
\end{itemize}
The second item needs a little more comment. The simplicity of the
data in this analysis is not completely accidental. Physicists tend to
include only those variables in analysis that obviously contribute to
the signal-background separation. The traditional approach is to
visually inspect one-dimensional distributions of a certain variable
for the signal and background components and include this variable
only if the two distributions are clearly distinct. With this mindset,
one will rarely find himself in a situation where a sophisticated
flexible classifier shows a significant improvement over simpler
methods, e.g., orthogonal cuts. Physicists are beginning to adopt more
advanced methods that search for trends, hardly visible to the human
eye, in data with dozens of input variables. In such analyses,
AdaBoost, decision trees and other powerful classifiers will prove
most useful.

Although none of the attempted classifiers gives a significantly
better performance over the original method designed by the analysts,
it is possible to improve the signal significance in this analysis by
a considerable amount. This can be accomplished by using a powerful
technique known in the statistics literature as ``bagging''. This
method will be described in a separate note.

Among the listed classifiers, AdaBoost with binary splits, the
decision tree and the bump hunter are the fastest. Training these
classifiers on a sample of 550k events takes only several minutes on a
Linux node with a 1.8~GHz CPU processor and 2~Gb of available
memory. Training of the AdaBoost classifier built with 50 decision
trees and the AdaBoost-based combiner of background-categorized
subclassifiers requires batch job submissions and consumes 4-8 hours
of running on allocated batch nodes. Because the CPU speed and
available memory vary from one batch node to another, a
straightforward comparison of the training time is not feasible. The
original method developed by the analysts also requires several hours
of training time; however, this training procedure is performed in a
substantially different framework and its speed could be significantly
limited by ROOT output.

\section{Summary}

A C++ package tailored for needs of HEP analysts has been implemented
and is available for free distribution to HEP researchers.

\begin{acknowledgments}
Thanks to Gregory Dubois-Felsmann for useful discussions and the idea
to sort terminal nodes of a decision tree by signal purity. Thanks to
Ed Chen for data and documentation on the $B\to\gamma l\nu$
analysis. Thanks to Byron Roe and Frank Porter for their comments on a
draft of this note.
\end{acknowledgments}

%\newpage

\end{document}